\documentclass[journal]{IEEEtran}

\usepackage{bm}
\usepackage{braket}
\usepackage{amsmath}
\usepackage{amssymb}
\usepackage{amsthm}
\usepackage{cite}
\usepackage{graphicx}
\usepackage{float}
\usepackage{algorithm}
\usepackage{algpseudocode}
\usepackage{xcolor}
\usepackage{url}

\usepackage{pgfplots}
\usepgfplotslibrary{groupplots}
\pgfplotsset{compat=1.18}

\newtheorem{theorem}{Theorem}

\theoremstyle{remark}

\begin{document}

\title{Beyond Typical Strings: Sampling the Complexity Landscape}

\author{Abdurrahman Hadi Erturk, Joshua Sharkey, Nora S. Martin,
Kamal Dingle, and Ard A. Louis%
\thanks{The work of A. H. Erturk was supported by the Clarendon Fund, University of Oxford. The work of N. S. Martin was supported by the Spanish Ministry of Science and Innovation through the Centro de Excelencia Severo Ochoa (CEX2020-001049-S, MCIN/AEI/10.13039/501100011033), the EMBL partnership and the Generalitat de Catalunya through the CERCA programme, and also the Issachar Fund. Research for this publication has been partially carried out in the Barcelona Collaboratorium for Modelling and Predictive Biology. This research is part of Grant JDC2022-049526-I funded by MCIN/AEI/10.13039/501100011033 and by European Union NextGenerationEU/PRTR. The funders played no role in the study design, data collection and analysis, decision to publish, or preparation of the manuscript.}
\thanks{A. H. Erturk and A. A. Louis are with the Department of Theoretical Physics, University of Oxford, Oxford, OX1 3PU, UK (e-mails: abdurrahman.erturk@physics.ox.ac.uk, ard.louis@physics.ox.ac.uk). N. S. Martin and J.Sharkey were also at this institution during part of this work.}%
\thanks{J. Sharkey is with the Institute for Gravitational Research, School of Physics and Astronomy, University of Glasgow, G12 8QQ, Glasgow, UK.}%
\thanks{N. S. Martin is with CRG (Barcelona Collaboratorium for Modelling and Predictive Biology), Dr. Aiguader 88, Barcelona, 08003, Spain.}%
\thanks{K. Dingle is with the Department of Mathematics and Natural Science,  Gulf University for Science and Technology, Kuwait.}%
\thanks{Ard A. Louis is the corresponding author}%
}

\markboth{}{Erturk \MakeLowercase{\textit{et al.}}: Beyond Typical Strings}

\maketitle

\begin{abstract}
Uniform random sampling of strings is overwhelmingly dominated by high-complexity strings, leaving the much smaller set of low- and intermediate-complexity strings largely unexplored. Here, we adapt the nested neutral set estimator (NNSE), a Monte Carlo rare-event sampling method, to estimate the distribution P(K) of complexity values for binary strings of fixed length. NNSE allows us to explore the full complexity landscape, including exponentially rare low-complexity regions that are inaccessible to direct sampling, for string lengths up to several hundred. We apply the method to several widely used complexity measures: Lempel–Ziv 76 (LZ76), Lempel–Ziv 78 (LZ78), context tree weighting (CTW), the block decomposition method (BDM), and block entropy. We further introduce uniform complexity weighting (UCW), which gives equal weight to different complexity levels rather than to individual strings. Using NNSE to perform UCW sampling allows us to compare complexity measures across their full range in a dataset-independent way. This reveals differences in their behavior for low- and intermediate-complexity strings that are practically inaccessible to conventional uniform sampling.
\end{abstract}


\IEEEpeerreviewmaketitle

\section{Introduction}

\IEEEPARstart{G}{iven} a complexity measure $K(x)$, estimated for example using code lengths from a compression algorithm, we consider the probability distribution $P(K)$ that one obtains complexity $K(x)$ upon picking a binary string $x$ at random from the whole set $\{0,1\}^n$ of strings of length $n$. For some measures, such as LZ78~\cite{ziv1978compression}, this distribution can be computed analytically~\cite{doganaksoy2006lempel}. In general, however, this distribution is not known.

Compression algorithms  are typically applied to natural datasets such as text, images, or biological sequences, which are much lower in complexity than typical random strings of the same length.   Many natural input–output maps are intrinsically biased towards producing low- to intermediate-complexity outputs~\cite{dingle2018input,dingle2020generic,mingard2019neural,johnston2022symmetry,mingard2025deep}. 
Understanding how compression-based complexity measures behave in the low and intermediate complexity regime, which forms a tiny  subset of sequence space,  is therefore of considerable practical interest.

One expects that, for many compression-based measures,  $P(K) \leq 2^{K}$~\cite{li2008introduction}. That is, low-complexity strings are exponentially rare under uniform random sampling.  As a result, comparisons of complexity measures have generally relied either on benchmark datasets, which introduce unknown source biases, or on uniform random sampling, which is dominated by high-complexity strings. There is a  need for a method that allows one to compare these measures on low to intermediate complexity strings that is also dataset agnostic.

Here, we introduce a nested Monte Carlo rare-event method inspired by work~\cite{jorg2008neutral} that estimates the sizes of neutral sets, defined as the sets of sequences mapping to the same RNA secondary structure. 
We adapt this nested neutral set estimator (NNSE) algorithm to binary strings and apply it to a series of  complexity measures. The NNSE samples uniformly within each complexity level across the whole range of $K$. 
This allows us to estimate the full distribution $P(K)$, including tail regions inaccessible to direct sampling.  It also provides a dataset-independent way to compare different complexity measures across the full range of $K$, rather than only for the typical high-complexity strings that dominate uniform random sampling.

More generally, it allows us to compare two distinct ways to measure properties of strings. We call the first, where all strings are treated as equally likely, \textit{uniform string weighting} (USW). The USW average of a function $f(x)$ of string $x$ over all $2^n$ strings of length $n$ is:
\begin{equation}
 \mathbb E_\text{USW} f(x) = 2^{-n} \sum_{\{0,1\}^n} f(x) 
\end{equation}
Many calculations in information theory use USW. However, this is dominated by strings with high complexity. 

An alternative way of sampling is \textit{uniform complexity weighting} (UCW) which gives equal weight to each complexity. For example, a UCW average would be:
\begin{equation}
    \mathbb E_\text{UCW} f(x) = \frac{1}{m} \sum_{K' = K_1}^{K_m} \frac1{N_{K'}} \sum_{x \in \{0, 1\}^n;\newline K(x) = K'} f(x)
\end{equation}
where $K: \{0, 1\}^n \to \{K_1, K_2, \dots, K_m\}$ where $K_1 < K_2 < \dots < K_m$ and $N_{K'} = |\{x \in \{0, 1\}^n: K(x) = K'\}|$. 
The NNSE allows us to  calculate UCW  properties of interest.  These would normally only be accessible through  exhaustive enumeration, which  is typically limited to string lengths of $n\approx30\!-\!40$, whereas our method extends this range to strings of length several hundred or longer.

The complexity distribution $P(K)$ is of interest beyond the comparison of complexity measures. The connection between lossless compression and universal prediction is well established~\cite{merhav1998universal}: the code lengths assigned by a universal compressor define an implicit probability distribution, $Q(x)\propto 2^{-K(x)}$, whose predictive performance depends on the distribution of $K$. A related connection arises in security, where randomized guessing strategies based on universal compression achieve optimal guessing exponents for decentralized brute-force attacks~\cite{merhav2019universal}. Here, $Q(x)$ determines the order in which candidate passwords are guessed. These applications provide further motivation for efficient methods to estimate $P(K)$, or equivalently to generate strings in the exponentially rare low-complexity regime where direct USW is ineffective.

\subsection*{Original Contributions}

\begin{itemize}
    \item We demonstrate that the NNSE gives accurate cumulative distributions $P(K)$ for five complexity measures on binary strings: LZ76~\cite{lempel1976complexity}, LZ78~\cite{ziv1978compression}, CTW~\cite{willems1995context, willems1998context},  BDM~\cite{zenil2016decomposition}, and block-entropy~\cite{cover2006elements}, reaching tail regions that are not accessible by direct sampling. 
    \item  We compare complexity measures using the NNSE to perform UCW sampling of strings under one measure $K_1(x)$ and evaluating those same strings under a second measure $K_2(x)$. This allows us to probe how well $K_1$ and $K_2$ correlate over the full range of complexity values.
    \item We also use  UCW sampling to study other properties of complexity measures.  For example, we show that calculating the  sensitivity of complexity to bit flips under UCW sampling can highlight features that are hard to distinguish under USW sampling. 
    \item We prove that the NNSE is ergodic for binary strings when a small mutation skip probability is introduced to the original algorithm~\cite{jorg2008neutral}. 
\end{itemize}

\section{Complexity Measures}

We consider the following complexity measures:  LZ76~\cite{lempel1976complexity}, LZ78~\cite{ziv1978compression}, CTW~\cite{willems1995context, willems1998context},  block entropy~\cite{cover2006elements}, and BDM~\cite{zenil2016decomposition}. 
We briefly summarize each below; see Appendix~\ref{app:complexity_measures} for details.

LZ76 is a classical parsing-based measure of finite-sequence complexity ~\cite{kaspar1987easily}, while LZ78 is a related dictionary-based universal coding scheme whose codelength provides a natural compression-based complexity measure. In contrast, context tree weighting (CTW) predicts each successive bit using a Bayesian mixture over context-tree models. We define the CTW complexity of a sequence as the cumulative log-loss incurred by these predictions. CTW has demonstrated strong predictive performance on real-world datasets~\cite{begleiter2004prediction} and is supported by a rigorous Bayesian framework for combining variable-order Markov models. 

Block entropy is particularly simple to understand. It calculates the complexity of a string using the Shannon entropy of fixed-length substrings (blocks). We use a sliding window with overlapping blocks to obtain a finer-grained estimate. Finally, we consider BDM, which approximates Kolmogorov complexity using precomputed algorithmic complexities of $n=12$ strings on 5-state Turing machines, which are then combined into blocks for a complexity measure. We have a separate subsection for BDM as both its uncorrelated (default) and correlated (slower but finer-grained) variants exhibit some unusual properties.

Many other complexity measures exist.  We chose the measures above because they are well established and span a range of approaches to practical string complexity. But NNSE can be applied directly to these and other measures.

\section{Methods}

\subsection{Nested Neutral Set Estimator (NNSE) for Strings}

The NNSE method, which we adapt from Jörg et al.~\cite{jorg2008neutral}, uses random walks on a population of strings $(x_0, \dots, x_m)$ to sample uniformly from small subsets of a larger parent set. It builds on the Metropolis-Hastings algorithm but differs in two ways: (1) instead of updating a single point $x$, we maintain a population $(x_0, \dots, x_m)$ at complexity levels $K_0, \dots, K_m$; and (2) we do not update each point in isolation but also swap neighbors $x_i, x_{i+1}$ when possible, sweeping  in increasing $i$. Like other nested sampling methods~\cite{skilling2006nested}, we subdivide the space into ordered levels, but our aim is to estimate the sizes of these nested sets rather than to evaluate an integral over likelihood values. We give pseudo-code in Algorithm 1 and 2. 

\algrenewcommand{\algorithmicindent}{0.8em}

\begin{algorithm}[t]
\caption{Nested Neutral Set Estimator (NNSE)}
\label{alg:nnse}
\begin{algorithmic}[1]
\small
\Require $K$; thresholds $K_0<\cdots<K_{m-1}$, $K_m=\infty$;
\Statex \hspace{\algorithmicindent} mutation op.\ $\mathcal{M}$; thermalization $T_0$; samples $T$; mutation-skip-probability $\eta$
\Ensure $\widehat{\pi}_i \approx |V_i|/|\mathcal{X}|$, histogram $\widehat{p}_i$
\State $x_m \gets \textsc{Uniform}(\{0,1\}^n)$
\State $x_i \gets \varnothing$,\ $h_i \gets 0$ \textbf{ for } $i<m$
\For{$t \gets 1$ \textbf{ to } $T_0+T$}
    \ForAll{$i$ with $x_i \neq \varnothing$} \Comment{parallel}
        \If{$\mathrm{Uniform}(0,1) > \eta$} \Comment{lazy step}
            \State $x' \gets \mathcal{M}(x_i)$ 
            \If{$K(x') \le K_i$} $x_i \gets x'$ \Comment{stay in $V_i$}
            \EndIf
        \EndIf
    \EndFor
    \State \Call{ExchangeSweep}{$x_0,\dots,x_m$}
    \If{$t > T_0$} \Comment{post-thermalization}
        \ForAll{$i\ge 1$ with $x_i \neq \varnothing$}
            \State $h_{i-1} \mathrel{+}= \mathbf{1}[K(x_i)\le K_{i-1}]$
        \EndFor
    \EndIf
\EndFor
\State $\widehat{r}_i \gets h_i/T$ \textbf{ for } $i<m$
\State $\widehat{\pi}_m \gets 1$;\;
       $\widehat{\pi}_i \gets \prod_{j=i}^{m-1}\widehat{r}_j$ \textbf{ for } $i<m$
\State $\widehat{p}_0 \gets \widehat{\pi}_0$;\;
       $\widehat{p}_i \gets \widehat{\pi}_i-\widehat{\pi}_{i-1}$ \textbf{ for } $i\ge 1$
\State \Return $(\widehat{\pi}_i)_{i=0}^{m},\ (\widehat{p}_i)_{i=0}^{m}$
\end{algorithmic}
\end{algorithm}

\begin{algorithm}[t]
\caption{Exchange sweep (increasing $i$)}
\label{alg:exchange}
\begin{algorithmic}[1]
\small
\Procedure{ExchangeSweep}{$x_0,\dots,x_m$}
    \For{$i \gets 0$ \textbf{ to } $m-1$}
        \If{$K(x_{i+1}) \le K_i$} \Comment{$x_{i+1}\in V_i$}
            \If{$x_i = \varnothing$} $x_i \gets x_{i+1}$
            \Else\ $x_i \leftrightarrow x_{i+1}$
            \EndIf
        \EndIf
    \EndFor
\EndProcedure
\end{algorithmic} \label{algorithm}
\end{algorithm}

Choosing closely spaced thresholds $K_i$ decomposes the string space $\mathcal X = \{0, 1\}^n$ into a nested family $V_0 \subset \dots \subset V_m = \mathcal X$, where $V_i = \{x \in \mathcal X : K(x) \leq K_i\}$ and $K_m$ is large enough that $V_m = \mathcal X$. The number of strings in $V_i$ (for $0 \leq i \leq m-1$) is
\begin{equation}
|V_i| = |V_m| \prod_{j = i+1}^{m} \frac{|V_{j-1}|}{|V_{j}|}
\end{equation}
and we estimate each ratio $\frac{|V_{j-1}|}{|V_j|}$ as follows: At each time step we flip a random bit of $x_j$ and keep the result only if it stays in $V_j$. This operation is termed a mutation and written as $x' \gets \mathcal M(x_j)$. In the original NNSE algorithm, $x_i$ is mutated at every step, but, for reasons explained below, we introduce a mutation-skip-probability $0 \leq \eta < 1$, similar to a lazy random walk. With probability $\eta$, we set $x' \gets x_i$ and we call this the lazy step. When the string $x'$ also lies in $V_{j-1}$, we record a \emph{hit} at $j-1$ and estimate
\begin{equation}
\frac{|V_{j-1}|}{|V_{j}|} \simeq \frac{h_{j-1}}{T}
\end{equation}
where $h_{j-1}$ is the number of hits at $j-1$ over $T$ iterations. We always use a single bit flip as the mutation operator but other mutation operators are also possible. 

The exchange sweep (Algorithm~\ref{alg:exchange}) additionally swaps $x_{j-1}$ and $x_j$ on a hit, which keeps the walk from becoming trapped near $x_j$ and lets it reach components of $V_j$ that lie far apart. For instance, the two lowest-complexity strings, $0\dots0$ and $1\dots1$, are maximally separated in Hamming distance; mutating $x_1$ alone could never reach both, whereas swapping lets us explore any component of $V_i$. 

Unlike Jörg et al.~\cite{jorg2008neutral}, we initialize only $x_m$ with a random point rather than seeding every $x_i$. Because the NNSE can already descend from $V_i$ to $V_{i-1}$, the remaining levels fill in on their own and this choice does not affect the final results. We discard a fixed time budget $T_0$ to let the population thermalize before recording hit counts.

The introduction of a lazy step is required for our proof of ergodicity in Appendix~\ref{app:analysis}. In practice, however, for the problems studied here we find that taking $\eta$ close to, or equal to, $0$ (i.e.\ removing the lazy step) does not noticeably affect the convergence of NNSE, and it is also faster.

Our proof of ergodicity is based on the existence of a reference state $(x_0,x_1,\dots,x_m)\in V_0\times\cdots\times V_m$ that can be reached from any other state, and from which any other state can in turn be reached. This construction allows us  to show that the Markov chain has a finite mixing time. The Markov-chain ergodic theorem~\cite{aldous-fill-2014} then guarantees that the NNSE estimates of the sizes of the sets $V_i$ converge almost surely to their true values as $t\to\infty$. Moreover, a Markov-chain Chernoff–Hoeffding bound~\cite{chung2012chernoff} allows us to relate the convergence time to the mixing time. Further details are given in Appendix~\ref{app:analysis}.

\subsection{Calculating $P(K)$ distributions}

After estimating the sizes of $V_i = \{x: K(x) \leq K_i\}$, we determine how many strings there are with complexity $K(x) \in (K_{i-1}, K_i]$ by calculating $|V_i \backslash V_{i-1}| = |V_i| - |V_{i-1}|$. This gives us the number of strings at a particular complexity value or a continuous interval of values, allowing us to calculate $P(K)$. 

For any complexity measure, we first determine the minimum and maximum complexities (or an estimate when unknown).  For example, for  LZ76, the min and max complexities are set at $\{k \times \log_2 n : k \geq 2 \text{ and } k \times \log_2 n \leq 1.3 \times n\}$. The minimum LZ76 complexity, attained by just repeating the same digit, is $2 \times \log_2 n$ and the maximum complexity is not known exactly but it is known that it decays to $n$ with increasing length \cite{estevez2013nonrandomness}. We use a factor of $1.3$ as a buffer to include the highest complexity strings.

For continuous complexity measures such as CTW, BE, and BDM, we discretize the complexity axis into a uniform grid of thresholds and use the resulting histograms to plot $P(K)$. CTW also has a maximum depth parameter $D$ which we set to $20$ for computational efficiency, as our tests show that CTW values vary very little for strings of length $n = 200$ above $D \geq 20$. For strings of length $n = 200$, CTW ($D = 20$) uses $419$ bins of width $\Delta K = 0.5$ on $[1, 210]$; BE uses $53$ bins of width $\Delta K = \log_2 n / 2$ on $[1, n]$; and BDM uses $200$ equally spaced bins on $[29.61, 577.04]$. 

In Appendix~\ref{app:validation} we confirm that the NNSE method recovers the exact $P(K)$ distribution of nested-set sizes for LZ76, LZ78, and CTW for exhaustive enumeration of all $n=30$ strings.

\subsection{Uniform Complexity Weighting (UCW)}

The NNSE can also be used to sample individual strings at prescribed complexity levels,  making it possible to perform  \emph{uniform complexity weighting} (UCW), in which strings are sampled with probability
\begin{equation}
P_{\mathrm{UCW}}(x)=\frac{1}{m}\frac{1}{N_{K(x)}},
\qquad
N_{K(x)}=\left|{y:K(y)=K(x)}\right|,
\end{equation}
where $m$ is the number of complexity levels for a discrete-valued measure $K$. For continuous measures, we first bin the complexity values into intervals $(K_0, K_1], \dots, (K_{m-1}, K_m]$. Under UCW, every complexity level has equal total probability, while strings within a level are sampled uniformly.  

Implementing UCW requires collecting strings throughout an NNSE run. Recording every visited string is impractical, since runs can be extremely long,  while sampling from a short time window would produce highly correlated samples. Instead, we use reservoir sampling~\cite{vitter1985random}. The first $c$ strings are stored, and each subsequent string encountered at time step $t$ replaces a uniformly chosen stored string with probability $c/t$. This produces a uniform sample of size $c$ over the entire run using only $O(c)$ memory, and at every stage remains uniform over all strings visited up to that point.

UCW sampling can be compared against the probability assignment $P_K$ that a complexity measure $K$ implicitly encodes, provided $K(x)$ can be interpreted as a description length or a code length, which is true for all the complexity measures we examine. By Shannon's source coding theorem, the optimal code lengths for a source $p$ are $\ell(x) = -\log_2 p(x)$~\cite{cover2006elements}; inverting this correspondence, the code lengths $K(x)$ are optimal precisely for a source assigning probability $P_K(x)$ proportional to $2^{-K(x)}$. Moreover, $\sum_x 2^{-K(x)} \leq 1$ in both cases: for a prefix-free compression algorithm this is Kraft's inequality~\cite{cover2006elements}, and for a probability model it follows from $q$ being a distribution. Hence, 
\begin{equation}
    P_K(x) = \frac{2^{-K(x)}}{Z}, \qquad Z = \sum_x 2^{-K(x)},
\end{equation}
is a well-defined distribution over binary strings. It is the source for which the code lengths $K$ would be optimal, i.e.\ the distribution implicitly assumed by the complexity measure.

The comparison of $P_K$ with UCW sampling is straightforward. Since $2^{-K(x)}$ is constant on each complexity level, both $P_K$ and $P_\text{UCW}$ are uniform within levels and differ only in the mass assigned across levels: level $k$ receives mass $1/m$ under $P_\text{UCW}$ and $N_k 2^{-k}/Z$ under $P_K$. Consequently,
\begin{equation}
    \frac{P_\text{UCW}(x)}{P_K(x)} = \frac{Z\, 2^{K(x)}}{m\, N_{K(x)}},
\end{equation}
and the two distributions coincide exactly when $N_k \propto 2^{k}$.

\subsection{Comparing Complexity Measures} \label{comparing_K_section}

Using the NNSE, we can compare complexity measures in a dataset-agnostic way. We perform UCW sampling with respect to one measure $K_1$ and evaluate the sampled strings under a second measure $K_2$, probing the relationship between the two measures across the full range of complexity values. The resulting samples estimate the conditional distribution
\begin{equation}
\begin{split}
    &P(K_2 = k_2 \mid K_1 = k_1) \\
    &= \frac{|\{x : K_1(x) = k_1,\ K_2(x) = k_2\}|}{|\{x : K_1(x) = k_1\}|},
\end{split}
\end{equation}
which describes how strings that $K_1$ regards as equally complex are distributed under $K_2$. In practice we visualize these conditionals as a scatter plot of $K_1(x)$ against $K_2(x)$ over the UCW samples and vice-versa. Comparing these two plots can give detailed insight into how complexity measures perform.

\pgfmathsetseed{42}

\newcommand{\uppertriscatter}[1][50]{%
  \foreach \i in {1,...,#1}{%
    \pgfmathsetmacro\ua{0.6 + 8.8*rnd}
    \pgfmathsetmacro\ub{0.6 + 8.8*rnd}%
    \pgfmathsetmacro\px{min(\ua,\ub)}%
    \pgfmathsetmacro\py{max(\ua,\ub)}%
    \edef\uppertripoint{\noexpand\fill[black] (axis cs:\px,\py) circle[radius=1.1pt];}%
    \uppertripoint
  }%
}

\pgfplotsset{
  noisy scatter/.style={
    only marks, mark=*, mark size=1.1pt, black,
    samples=15, domain=0.6:9.4,
  },
  K subplot/.style={
    width=3.6cm, height=3.6cm,
    xmin=0, xmax=10, ymin=0, ymax=10,
    axis lines=left,
    ticks = none,
    font=\footnotesize,
    xlabel shift=3pt, ylabel shift=3pt,
  },
}
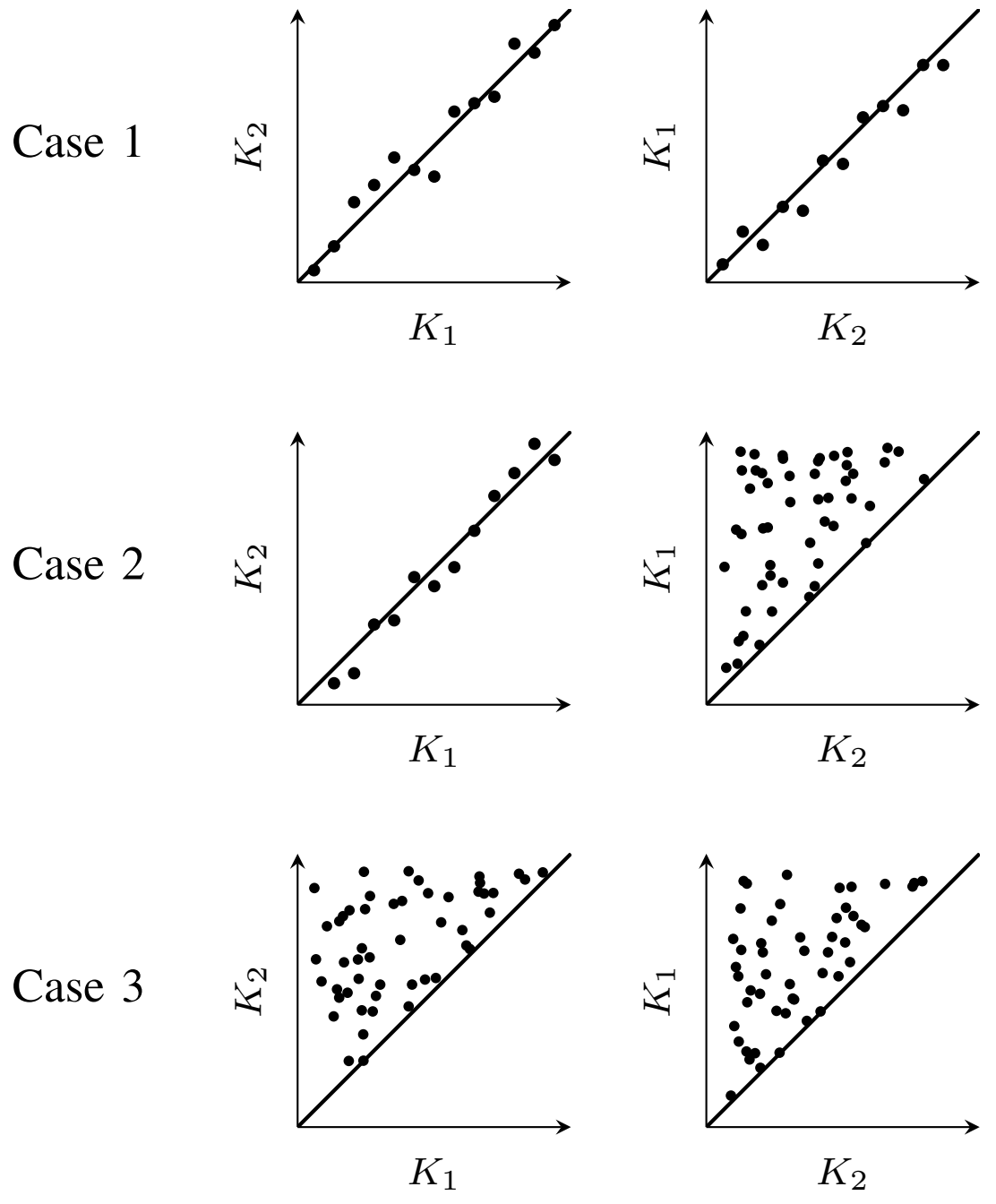
\begin{figure}[ht]
  \centering
  \begin{tikzpicture}
    \begin{groupplot}[
      group style={
        group size=2 by 3,
        horizontal sep=1.0cm,
        vertical sep=1.1cm,
      },
      K subplot,
    ]
    \nextgroupplot[xlabel={$K_1$}, ylabel={$K_2$}]
    \addplot[black, thick, domain=0:10] {x};
    \addplot[noisy scatter, samples=13] {x + 0.6*(rand + rand + rand)};

    \nextgroupplot[xlabel={$K_2$}, ylabel={$K_1$}]
    \addplot[black, thick, domain=0:10] {x};
    \addplot[noisy scatter, samples=13] {x + 0.6*(rand + rand + rand)};

    \nextgroupplot[xlabel={$K_1$}, ylabel={$K_2$}]
    \addplot[black, thick, domain=0:10] {x};
    \addplot[noisy scatter, samples=13] {x + 0.6*(rand + rand + rand)};

    \nextgroupplot[xlabel={$K_2$}, ylabel={$K_1$}]
    \addplot[black, thick, domain=0:10] {x};
    \uppertriscatter

    \nextgroupplot[xlabel={$K_1$}, ylabel={$K_2$}]
    \addplot[black, thick, domain=0:10] {x};
    \uppertriscatter

    \nextgroupplot[xlabel={$K_2$}, ylabel={$K_1$}]
    \addplot[black, thick, domain=0:10] {x};
    \uppertriscatter

    \end{groupplot}
    \node[anchor=south]
      at ([xshift=-4.6em, yshift=-0.6em]group c1r1.west) {Case 1};
    \node[anchor=south]
      at ([xshift=-4.6em, yshift=-0.6em]group c1r2.west) {Case 2};
    \node[anchor=south]
      at ([xshift=-4.6em, yshift=-0.6em]group c1r3.west) {Case 3};
  \end{tikzpicture}
  \caption{Illustrating three idealized cases (rows 1-3) when  comparing complexity measures $K_1$ and $K_2$.  Left column plots correspond to strings sampled UCW with respect to $K_1$ and right column plots are the opposite order.}
  \label{fig:k1k2}
\end{figure}

We will treat three simple idealized cases to illustrate what kind of information we can derive from these subplots: 
\begin{itemize}
    \item \emph{Both straight.} (Top row of Figure~\ref{fig:k1k2}) Both plots of $P(K_2$$\mid$$K_1)$ and $P(K_1$$\mid$$K_2)$ show a linear correlation between $K_1$ and $K_2$, suggesting that both measures estimate complexity similarly.

    \item \emph{One triangular, one straight.} (Middle row of Figure~\ref{fig:k1k2}) The plot of $P(K_2$$\mid$$K_1)$ has an upper triangular shape but that of $P(K_1$$\mid$$K_2)$ shows a linear relationship. This  indicates that there are many strings to which $K_1$ assigns a high complexity but $K_2$ assigns a low complexity. However, the reverse does not hold: most of the strings which $K_1$ assigns a low complexity to, $K_2$ also assigns a low complexity to. We can therefore infer that $K_2$ captures the patterns identified by $K_1$, while also exploiting additional regularities that $K_1$ do not. This suggests that $K_2$ is a more general complexity measure than $K_1$. 

    \item \emph{Both triangular}. (Bottom row of Figure~\ref{fig:k1k2}). Both plots of $P(K_2$$\mid$$K_1)$ and $P(K_1$$\mid$$K_2)$ have an upper triangular shape. This means that each measure exploits different kinds of patterns or regularities in strings.
\end{itemize}

Finally, it may appear paradoxical that the two plots in the middle row of Figure~\ref{fig:k1k2} disagree, since both are conditionals of the same joint distribution: the triangular plot exhibits many strings with low $K_1$ and high $K_2$, while the straight plot, sampled at exactly those high $K_2$-levels, shows no trace of them. The resolution is a counting argument. The level set $\{x : K_2(x) = k_2\}$ at high $k_2$ contains on the order of $2^{k_2}$ strings, the overwhelming majority of which have high complexity under $K_1$; the strings with anomalously low $K_1$ within this set are exponentially few by comparison. Uniform sampling within this level set therefore draws high $K_1$ complexity strings with probability exponentially close to one, and the anomalous low $K_1$ strings never appear. UCW sampling with respect to $K_1$, by contrast, allocates equal probability to each $K_1$-level, and thus low $K_1$ high $K_2$ strings can appear prominently. The two plots are thus consistent: they weight the same underlying set of strings differently, and their disagreement is precisely what reveals the differences between different complexity measures in the low-complexity regime, which is effectively inaccessible to uniform random sampling.

\section{Results}

\subsection{Complexity Distributions $P(K)$}

\begin{figure}
    \centering
    \includegraphics[width=1.0\linewidth]{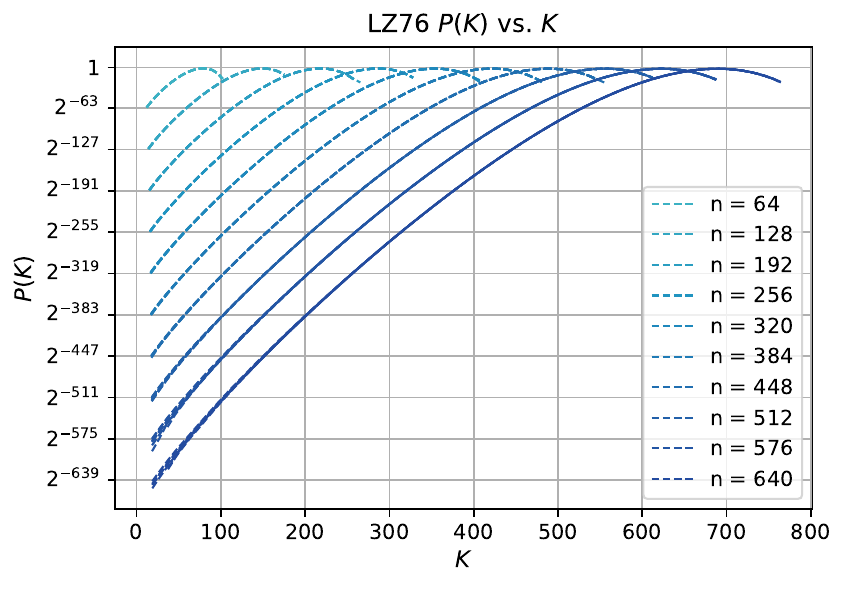}
    \caption{Complexity distribution $P(K)$  calculated using the NNSE method, using LZ76 complexity, for  string lengths $64 \leq n \leq 640$  with $\Delta n = 64$ increments. 
    The  horizontal lines at $y = 2^{-n+1}$, correspond to the probability of the two simplest strings for length $n$. For each length we show 8  independent runs to  help illustrate statistical variation between runs,  which becomes larger for longer lengths and at lower complexity. The curves look roughly like $P(K) \propto 2^K$, as expected.
    }
    \label{fig:LZ76_64_704_histogram}
\end{figure}

To illustrate our method, we calculated $P(K)$ for LZ76 using the NNSE for $10^7$ time steps (plus $10^6$ steps for thermalization) for string lengths ranging from $n=64$ to $n=640$. The plots are shown  in Figure \ref{fig:LZ76_64_704_histogram}. 
The LZ76 distributions scale smoothly with string length, and independent NNSE runs are almost indistinguishable, indicating small statistical errors. While this does not guarantee accuracy, it is encouraging, especially given the agreement with exact enumeration for $n=30$ (Appendix~B). A second consistency check is that the smallest probability, $P(K=K_{\min})$, is very close to $2^{-n+1}$ for every $n$. This is expected because there are exactly two strings with minimum LZ76 complexity: $00\ldots0$ and $11\ldots1$.
For all string lengths, the distribution exhibits a peak near $K=n$ with an approximately Gaussian shape, as predicted theoretically~\cite{jacquet1995asymptotic}. Over much of the remaining range, $\log_2 P(K)$ is nearly linear in $K$, consistent with the redundancy bounds relating LZ76 complexity to source entropy~\cite{lempel1976complexity}

\begin{figure}
    \centering
    \includegraphics[width=1.0\linewidth]{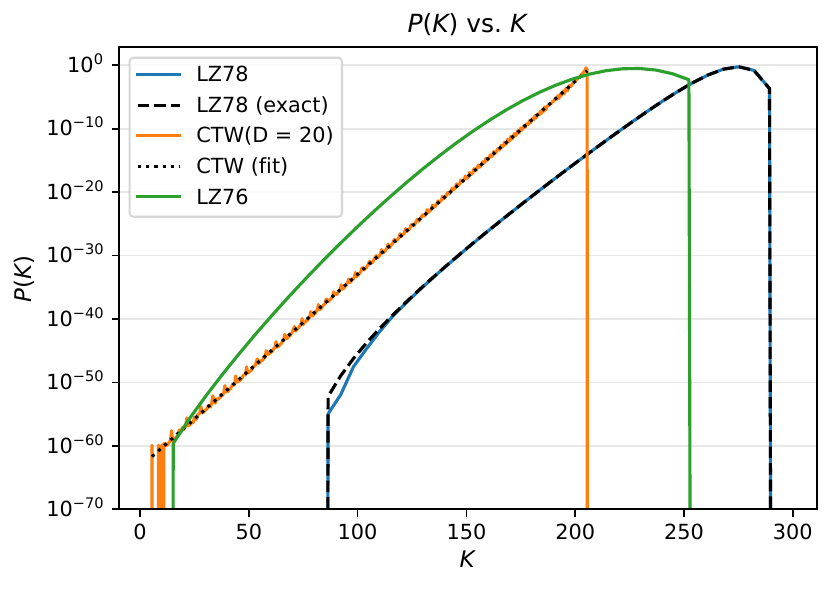}
    \caption{Complexity distributions $P(K)$ for binary strings of length $n = 200$, obtained with the NNSE method for the complexity measures LZ76, LZ78, and CTW $(D = 20)$. The dotted curve is an exponential approximation to the CTW distribution, $\log_2 P(K) \approx 1.01\,K - 210.5$. The dashed line is the exact LZ78 distribution.}
    \label{fig:n=200_logP}
\end{figure}

Figure~\ref{fig:n=200_logP} compares the complexity distributions of LZ76, LZ78, and CTW for strings of length $n=200$. We ran NNSE for $10^9$ iterations for LZ76 and LZ78, but only $10^7$ iterations for CTW because it converged much more rapidly. For efficiency, CTW was evaluated with context depth $D=20$ rather than the maximum possible depth $D=200$. Since our tests showed no noticeable difference between these choices, this does not materially affect the results. 

We note that LZ76 and LZ78 are qualitatively similar in shape, with a Gaussian-like shape near their maxima, and a more linear $\log_2 P$ vs.\ $K$ relationship at lower complexities. By contrast, the CTW plot is, on average, very close to linear on its entire range, much like Shannon entropy is. This average is in general quite well fit by a simple form $\log_2P(K) = a + bK $.  CTW also exhibits more complex spiking structure, especially at lower complexities.  As shown in Appendix~\ref{app:ctw},  this structure is stable under NNSE runs and holds for different lengths.  It can also be observed in the exact enumeration in Appendix~\ref{app:validation}. This feature of CTW is not, to our knowledge, yet understood theoretically. It illustrates how the NNSE can uncover new patterns in complexity measures. 

While we believe that the plots in Figure~\ref{fig:n=200_logP} for LZ76 and CTW are fully converged, this is not the case for LZ78.  We compare it to the exact distribution obtained via a polynomial-time algorithm that calculates the exact number of strings with a specific LZ78 complexity analytically \cite{doganaksoy2006lempel}. For most of the range the NNSE closely tracks the exact solution, but there are discrepancies at smaller $K$.  We trace the origin of these sampling difficulties to the much higher sensitivity of LZ78 to single bit flips, a property we study in more detail in the next section.  We also note that in Appendix~\ref{app:lz76_symmetric} the NNSE converges more rapidly for a symmetrized version of LZ76, than for the standard LZ76, and that this may be due to the former's lower sensitivity to single bit flips.

\subsection{USW and UCW calculations of  single-bit sensitivity}

  We define the single-bit sensitivity  $S_1^n(K_i)$ as the average change in the complexity assigned by measure $K_i$ when a single bit of a string of length $n$ is flipped. This average can be computed either by USW or UCW.  
Table~\ref{tab:sensitivity} reports normalized versions of  both estimates for the five complexity measures considered here, computed from $10{,}000$ strings of length $n=200$. We include uncorrelated BDM in the table, but not the correlated variant. The UCW estimates were obtained using NNSE. For all five measures, the UCW values exceed the corresponding USW values, in some cases by more than an order of magnitude. This indicates that low-complexity strings are, on average, substantially more sensitive to single-bit perturbations than typical strings.

Perhaps more strikingly, the complexity measure with the lowest USW sensitivity might not be the one with the lowest UCW sensitivity. Under USW sampling, the normalized $S_1^{200}(\mathrm{LZ78})$ is only slightly larger than $S_1^{200}(\mathrm{LZ76})$, while $S_1^{200}(\mathrm{BDM})$ is considerably smaller than $S_1^{200}(\mathrm{LZ76})$. Under UCW sampling, however, BDM has a larger normalized sensitivity than LZ76, and that of LZ78 is five times larger than that of LZ76. As we show below, this difference arises because LZ78 and BDM exhibit much greater single-bit sensitivity for low-complexity strings, a feature that is largely obscured under USW sampling.

\begin{figure}
    \centering
    \includegraphics[width=1.0\linewidth]{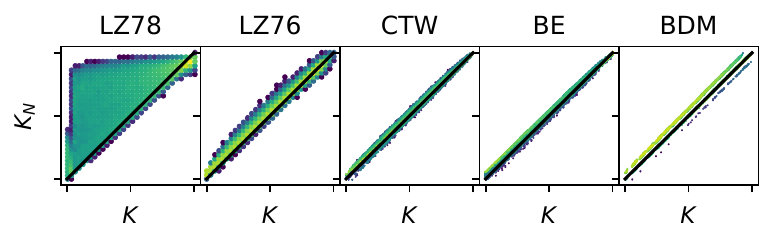}
    \caption{Scatter plots of complexity of a string upon a single bit-flip ($K_N$) vs.\ its original complexity $K$ for (a) LZ78, (b) LZ76, and (c) CTW($D = 20$), (d) BE($W = 10$), and (e) BDM for strings of length $n=200$. First, $100{,}000$ strings are sampled UCW from an NNSE run with 1 billion main steps. Then a single bit is flipped for all strings to obtain a neighboring string, one Hamming distance apart. The new complexity is plotted on the y-axis, and the original complexity on the x-axis.  The deviation from $x=y$ indicates how sensitive a complexity measure is to single-bit flips. Brightness indicates density in the figures.}
    \label{fig:n=200_neighbor_complexity}
\end{figure}

To illustrate what UCW captures that USW does not, Figure~\ref{fig:n=200_neighbor_complexity} shows, for each of the $100,000$ strings of length $n = 200$,  the complexity of a string after flipping a single bit, plotted against the original complexity of the string for LZ78, LZ76, and CTW. 
The first clear feature is that most bit flips increase complexity, especially for strings that are initially of low complexity. This is expected: low-complexity strings are atypical, and a random bit flip is therefore more likely to move them towards more typical strings.

Perhaps more strikingly,  LZ78 exhibits substantially larger changes in complexity following a single-bit flip than either LZ76 or CTW, with the effect being most pronounced at lower complexities. This helps explain  why the mean single-bit sensitivity measured by UCW is much higher for LZ78 than for the other two methods.  The large UCW sensitivity of LZ78 and BDM suggests that the patterns these measures identify are comparatively fragile, particularly in the low- and intermediate-complexity regime. This behavior is not apparent from USW alone.

Finally, the large sensitivity helps explain why the NNSE has more difficulty converging for LZ78 at low complexities; the sampling is more likely to get stuck in local minima.   By contrast, CTW is the least sensitive, and indeed, we find that the NNSE converges for much longer lengths, up to $n \approx 1000$ without too much difficulty.  

\begin{table}[t]
\caption{Mean single-bit sensitivity  $S_1^n(K_i)$ of complexity measures $K_i$ under USW and UCW sampling for strings of length $n=200$.  Values in brackets are normalized to the range  $[0,200]$}
\label{tab:sensitivity}
\centering
\setlength{\tabcolsep}{4pt}
\begin{tabular}{||c | c c c c c||}
 \hline
 Sampling & LZ78 & LZ76 & CTW & BE ($W{=}10$) & BDM \\ [0.5ex]
 \hline\hline
 USW & 3.0 (3.0) & 3.0 (2.5) & 0.18 (0.18) & 0.41 (0.53) & 1.5 (0.56) \\
 \hline
 UCW & 36 (35) & 7.8 (6.6) & 4.1 (4.1) & 3.1 (4.0) & 19 (7.2) \\
 \hline\hline
\end{tabular}
\end{table}

\subsection{Comparing complexity measures through cross-correlations}

\begin{figure}
    \centering
    \includegraphics[width=1.0\linewidth]{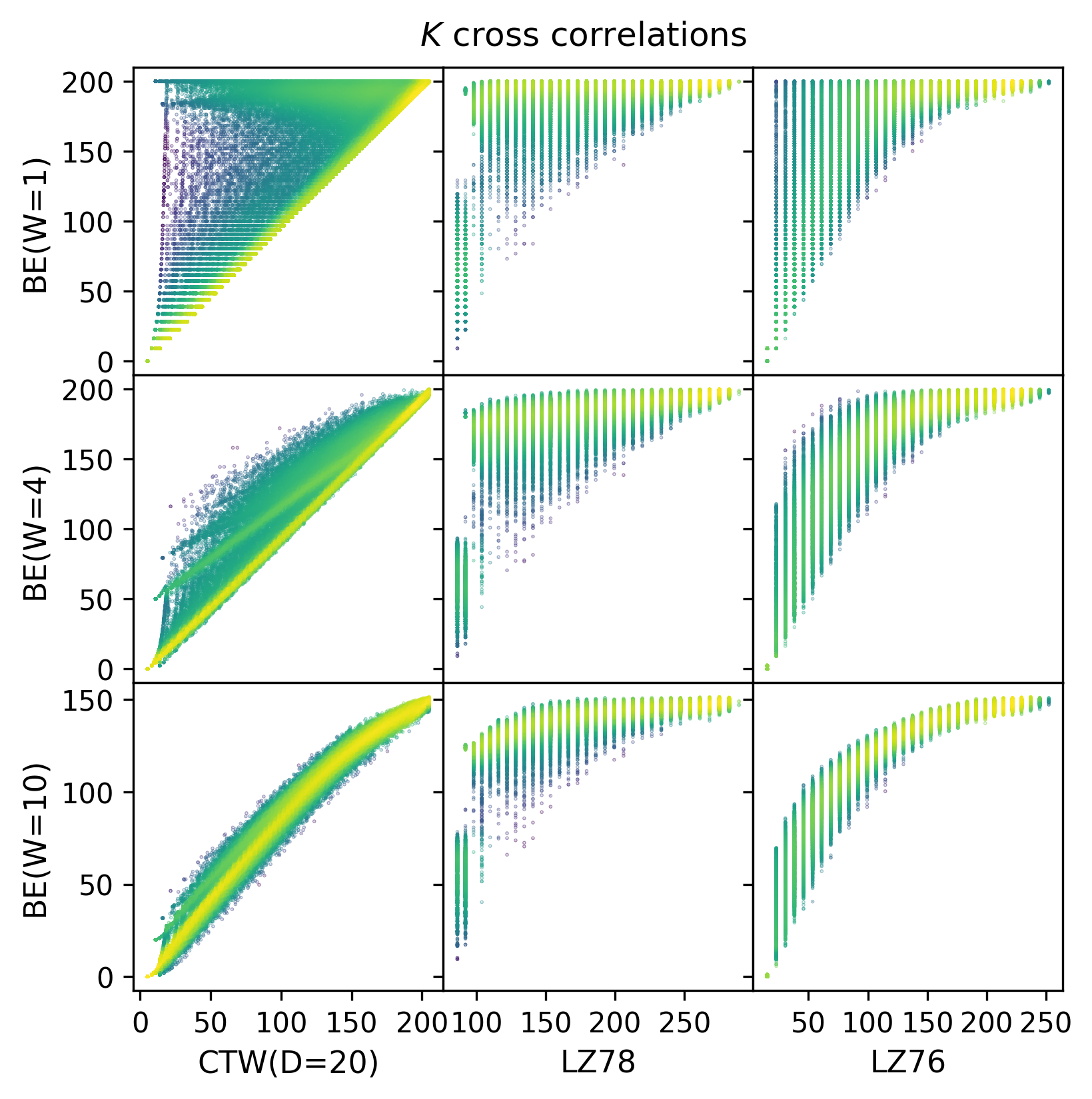}
    \caption{LZ76, LZ78, and CTW cross correlations with block entropy with different block widths for $100,000$ $n=200$ strings. 
     In each column, strings are sampled uniformly from successive complexity levels of the measure indicated at the bottom (left to right: CTW, LZ78, and LZ76).  The x-axis gives a string's complexity under the sampling measure, while the y-axis gives its complexity measured by block entropy. The color indicates density with brighter colors indicating a higher density of points. 
     The upper triangular shape in the top row arises because there are many low-complexity strings with high Shannon entropy, such as $0101...01$.
     }
    \label{fig:placeholder}
\end{figure}

\begin{figure}
    \centering
    \includegraphics[width=1.1\linewidth]{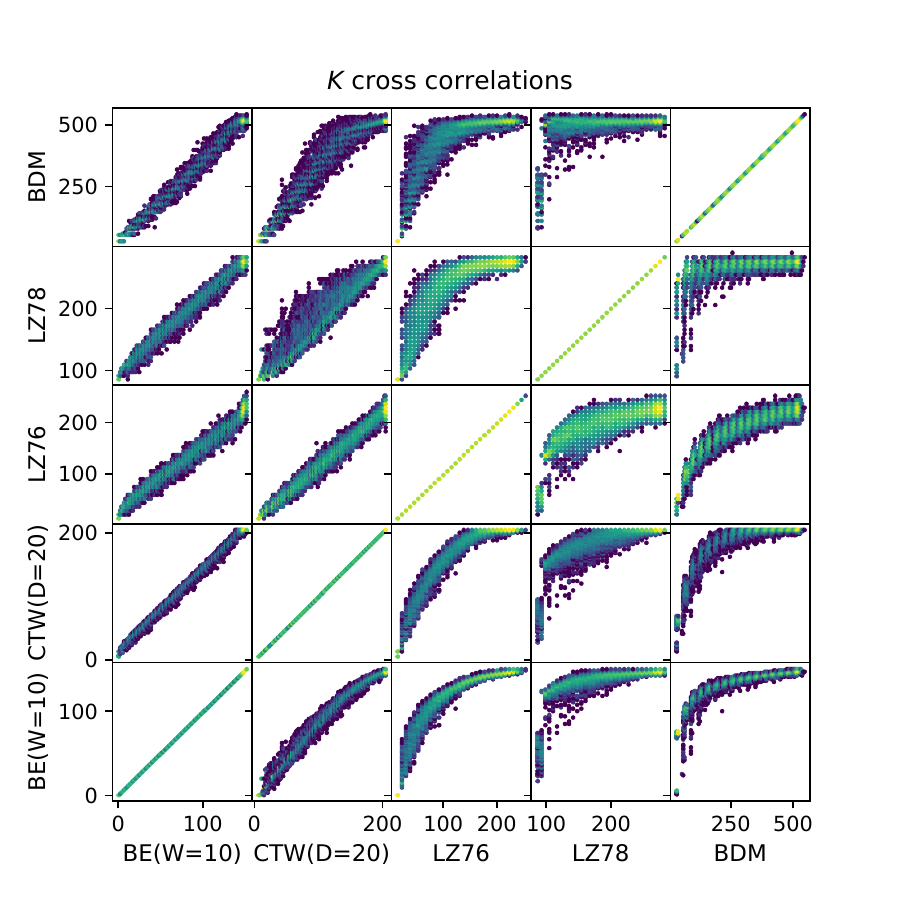}
    \caption{Scatter plots comparing  complexity measures BE(W = 10), LZ76, LZ78, CTW($D=20$), and BDM for $100,000$ strings of length $n=200$, obtained using the NNSE method. In each column, strings are sampled uniformly from successive complexity levels of the measure indicated at the bottom. Each point corresponds to a sampled string. The x-axis gives its complexity under the sampling measure, while the y-axis gives its complexity under the measure shown for that row. Brighter colors indicate regions with higher density. The complexity values of continuous complexity measures, CTW, BE, and BDM, have been discretized.}
    \label{fig:pairwise_plots}
\end{figure}

We next used the methods described in section~\ref{comparing_K_section} to cross-correlate LZ76, LZ78, CTW($D=20$), BE($W=10$), and BDM. We ran NNSE  for $10^9$ time steps (with $10^8$ thermalization steps) and sampled $100{,}000$ strings for each level. The sampled strings need not be unique. We plot the results in Figures \ref{fig:placeholder} and \ref{fig:pairwise_plots}. 

It is instructive to see how CTW, LZ78, and LZ76 correlate with different levels of block entropy in Figure.~\ref{fig:placeholder}.  There are many low complexity strings that have high Shannon entropy BE($W=1$).   It is already known that LZ76, LZ78, and CTW can compress sequences from an entropic source. However, we also demonstrate the reverse implication, that low complexities also tend to have low block entropies as the block size becomes larger.  Nevertheless,  in each case there are strings that these measures consider simple, but which the block entropies consider more complex.   We also note an interesting property of CTW, which shows one or more high intensity lines (especially visible with BE($W=4$)) in addition to the $x = y$ line.  Further examples of this pattern are shown in Appendix~\ref{app:ctw}, see especially Figure~\ref{fig:ctw_fine_grained}. 

We cross-correlate these measures in Figure~\ref{fig:pairwise_plots}. One striking result can be seen in the first column from a qualitative analysis.  All four more sophisticated measures correlate remarkably well with BE($W=10$). In other words, strings that have low BE($W=10$) block entropy tend to also have low complexities. CTW has the closest relationship to block entropy, which is explained by the fact that CTW does a Bayesian weighting over different context trees and each context tree has a KT prior on its prediction parameters, that closely mirrors Shannon entropy. Note that the gap in LZ78 strings with lower complexities is caused by the same sampling issues we identified earlier in our sensitivity analysis.

A qualitative comparison to the schematic three cases from Figure~\ref{fig:k1k2} shows that CTW and LZ76 are closest to case 1, except at high complexities where LZ76 complexities are more spread out across the Gaussian part of LZ76’s $P(K)$.   It is perhaps somewhat surprising that LZ76 is more closely correlated with CTW than LZ78, even though both LZ76 and LZ78 are deterministic parsing algorithms based on substring matches, whereas CTW is a Bayesian ensemble prediction method.  

Perhaps the closest example to case 2 is the comparison between LZ76 and BDM. Strings with low or high complexity according to BDM generally also have low or high complexity according to LZ76. The converse, however, holds less consistently: there are substantially more strings that LZ76 assigns low complexity but BDM assigns high complexity than vice versa.

LZ78 and BDM are the closest to our case 3: both correlations show upper-triangular form suggesting that these two compression methods uncover different patterns in strings. 

\subsection{Block Decomposition Method (BDM)}
\begin{figure}
    \centering
    \includegraphics[width=1.0\linewidth]{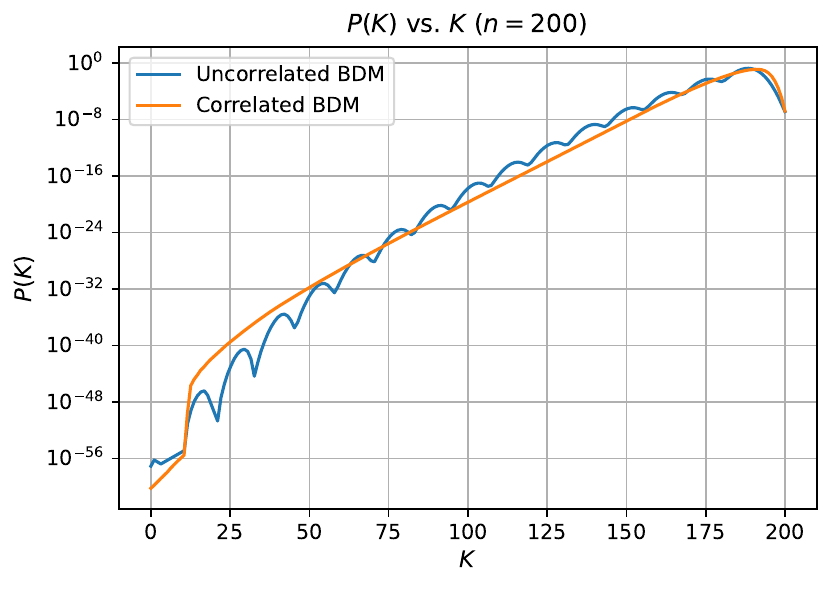}
    \caption{BDM complexity distribution $P(K)$ for $n=200$, for both the uncorrelated (blue) and correlated (orange) variants. The x-axis is normalized on $[0,200]$.}
    \label{fig:bdm}
\end{figure}

The final measure we study is BDM~\cite{zenil2014correlation}, because its use of short length $12$ strings with their complexity measured by a coding theorem method (CTM)~\cite{soler2014calculating} based on small universal Turing machines differs from other complexity measures. The complexity of the full dataset is  estimated by aggregating the CTM values of unique blocks according to the formula $\mathrm{BDM}(s) = \sum_{i} \mathrm{CTM}(s_i) + \log_2(n_i)$, where $s_i$ represents each unique block and $n_i$ its number of occurrences. 

We applied the NNSE method to two variants of BDM: the standard uncorrelated version and the correlated version, which uses a sliding window and is computationally more expensive. For the uncorrelated variant, we sampled 200 complexity values from the minimum of approximately $30$ to  the maximum at approximately $580$. The correlated variant spans a much wider range of complexities and so we instead sampled $200$ geometrically spaced values between $30$ and $6600$. Since the two methods produce complexity values on very different scales, we linearly rescaled the horizontal axis of each curve to the interval $0$–$200$.

For the standard uncorrelated BDM, we estimated $P(K)$ for binary strings of length $n=200$ using NNSE with $T=10^8$ sampling steps after $10^7$ thermalization steps. The resulting distribution is shown in Figure~\ref{fig:bdm}. A striking feature is the presence of pronounced oscillations in $\log_2 P(K)$.  We repeated the calculation for other string lengths and found the same pattern in every case. This suggests that the oscillations are an intrinsic feature of the uncorrelated BDM, most likely arising from its non-overlapping block decomposition, see Appendix~\ref{app:bdm}. 

We ran NNSE on the correlated BDM using as many steps as for the uncorrelated variant. Correlated BDM does not exhibit any oscillations but it does exhibit a sharp transition near $K\approx400$ (corresponding to approximately $12$ on the normalized axis in Figure~\ref{fig:bdm}), where the probability drops by about four orders of magnitude. On either side of this transition, $\log_2 P(K)$ is approximately linear in $K$, although the slopes differ slightly between the two regimes.  To the best of our knowledge, these unusual behaviors in the uncorrelated and correlated version of BDM have not been reported previously.

\section{Discussion}
The principal contribution of this work is methodological. We show that the NNSE provides practical access to exponentially rare regions of complexity space that are effectively inaccessible under uniform random sampling. This makes it possible to estimate the distribution $P(K)$ over many orders of magnitude and to study complexity measures across their full range of values rather than only in the neighborhood of typical high-complexity strings.   For binary strings, this extends the practical range of exhaustive analysis from lengths of roughly $n\approx30 $ to several hundred or  more, while remaining independent of any particular dataset.

While the NNSE can access the expected near exponential scaling of $P(K)$ at low $K$, it also  reveals unexpected results, such as  reproducible fine-scale structure in CTW and previously unreported oscillations in the default implementation of BDM, and a sharp transition at low complexity in the correlated variant.  

A second more general contribution is the introduction of uniform complexity weighting (UCW). Many calculations in  information theory use uniform string weighting (USW), where every string is equally likely. Although this is a natural choice, USW is overwhelmingly dominated by strings of near-maximal complexity. Consequently, it can obscure precisely the region of complexity space that is often of greatest practical interest.  UCW instead gives equal weight to each complexity level, allowing low- and intermediate-complexity strings to be studied without introducing biases associated with particular datasets. This change in viewpoint can lead to qualitatively different conclusions. For example, the average single-bit sensitivity of every complexity measure increases under UCW, with the largest changes occurring for LZ78 and BDM. These differences between measures  are not apparent under USW because low-complexity strings form an exponentially small fraction of sequence space.

UCW also provides a natural framework for comparing complexity measures directly, revealing patterns that would be much harder to observe using conventional sampling approaches.  Rather than asking whether two complexity measures perform similarly on a particular corpus, this approach asks whether they assign similar notions of complexity across the entire space of binary strings. For example, the cross-correlations suggest that LZ76 and CTW produce remarkably similar complexity orderings despite arising from rather different principles. 

The present work also has several limitations. Although NNSE reproduces exact complexity distributions whenever exhaustive enumeration is feasible, we do not yet have rigorous theoretical bounds on its convergence rate or sampling error. The behavior of LZ78 illustrates this limitation: its comparatively high single-bit sensitivity makes low-complexity regions more difficult to explore and leads to significantly slower convergence than for LZ76 or CTW. Likewise, several empirical observations reported here—including the spike structure of CTW and the oscillatory behavior of uncorrelated BDM—remain theoretically unexplained. These should therefore be regarded as observations generated by systematic computation rather than as fully understood mathematical phenomena.

Our framework is readily extensible. Since the NNSE depends only on the ability to evaluate a local measure, it can be applied to essentially any computable description-length estimator, including modern neural compression methods~\cite{balle2017endtoend}, probabilistic sequence models~\cite{mahoney2005adaptive}, and future approximations to Kolmogorov complexity. Beyond complexity measures, the same rare-event methodology may prove useful whenever one wishes to characterize exponentially small subsets of very large combinatorial spaces.

While this paper is primarily empirical and methodological, perhaps the most interesting longer-term implication concerns theory itself.   Many existing theoretical results concern asymptotic behavior under USW, whereas practical applications often depend on the comparatively small subset of low- and intermediate-complexity strings.  It would be valuable to develop analytical results directly for UCW, just as many existing results have been developed for USW.

\bigskip

\appendices

\section{Complexity Measures}\label{app:complexity_measures}

\subsection{Lempel-Ziv 1976}

This complexity measure was introduced by Abraham Lempel and Jacob Ziv~\cite{lempel1976complexity} and is commonly abbreviated as LZ76. It factorizes a binary string $x = x_1x_2\dots x_n$ into factors $f_1 \dots f_m = x$, where each $f_i$ is the shortest prefix of $f_i f_{i+1} \dots f_m$ that does not occur as a substring of $f_1 \dots f_i$ except as a suffix. Simply put, at each step, LZ76 finds the longest match with the text seen so far and extends it by one new symbol. The LZ76 complexity of a string is the number of factors $m$ multiplied by $\log n$. The product $m\log n$ approximates the description length of $x$ since each factor can be encoded using $O(\log n)$ bits. LZ76 complexity can be computed in $O(n\log n)$ time using suffix automata~\cite{blumer1985}. 

\subsection{Lempel-Ziv 1978}

Commonly abbreviated as LZ78, this is a dictionary-based compression method introduced by Ziv and Lempel~\cite{ziv1978compression}. LZ78 factorizes a binary string $x = x_1 \dots x_n$ into a sequence of factors $f_1 \dots f_m$ such that each factor $f_i$ is the shortest prefix of the remaining string that is distinct from all previous factors $f_j$ with $j < i$. Each factor thus extends a previous one by one bit. LZ78 complexity is given by $m \log m$ to approximate description length. 
LZ78 complexity can be computed in $O(n)$ time \cite{lz78_linear}. 

\subsection{Context Tree Weighting}

 Context tree weighting (CTW), first proposed by Willems et al.~\cite{willems1995context}, is a compression method that has good theoretical guarantees and strong practical performance. It calculates a conditional probability distribution $P_\text{CTW}(x_t|x_1 \dots x_{t-1})$ for the next bit while parsing the string, using a weighted mixture over context trees to model the string. We obtain the CTW complexity as the total log-loss of these predictions:
\begin{equation}
    K_\text{CTW}(x) = - \sum_t \log_2 P_\text{CTW}(x_t|x_1 \dots x_{t-1}).
\end{equation}
CTW has one parameter $D$ which determines the maximum depth of the context tree it uses to model the sequence. CTW runs in $O(nD)$ time \cite{veness2011monte} and is slower than the $LZ$ family of algorithms in practice. 

\subsection{Block Entropy}

Block entropy is another way to measure the complexity of strings~\cite{cover2006elements}. Shannon entropy for a general distribution $p: \mathcal X \rightarrow [0, 1]$ is defined as 

\begin{equation}
    H(p) = - \sum_{x \in \mathcal X} p(x) \log p(x)
\end{equation}

For binary strings, $\mathcal X = \{0, 1\}$, which limits the usefulness of Shannon entropy. However, we can define block entropy by letting $\mathcal X = \{0, 1\}^k$ for any $k$ where $k$ is called the \emph{block size}. We calculate block entropy as 
\begin{equation}\label{block_entropy}
    BE_k(x) = - \sum_{y \in \{0,1\}^k} P_x(y) \log_2 (P_x(y))
\end{equation}
where $P_x(y)$ counts the number of times a substring $y = y_1 \dots y_k$ occurs in the string x and divides that to the total number of length $k$ substrings in $x$, i.e.\ 
\begin{equation}\label{pxy}
    P_x(y) = \frac{|\{1 \leq i \leq n - k + 1: x_i \dots x_{i + k - 1} = y_1 \dots y_k\}|}{n - k + 1}
\end{equation}
When $k = 1$, this gives the Shannon entropy of a sequence. The description length of a string is given approximately as 
\begin{equation}
    K_{BE}(x) = \frac{n}{k} BE_k(x)
\end{equation}

Although choosing larger $k$ allows us to examine longer patterns in sequences, it also slows the convergence of block entropy to the true distribution. Since the size of $\mathcal X$ grows exponentially with $k$, we need a length of at least $2^k + k - 1$ to pack every possible substring in one string. Thus, computing block entropy for short strings will almost always underestimate the true entropy rate of the distribution, but this error vanishes in the infinite $n$ limit. 

\subsection{Block Decomposition Method}

The block decomposition method (BDM)~\cite{zenil2016decomposition} approximates the algorithmic complexity of a string by decomposing it into smaller blocks (up to size 12) for which complexity values have been precomputed via exhaustive enumeration of small Turing machines with 5 states and 2 symbols~\cite{soler2014calculating}. The complexity of the full string is then estimated by aggregating the CTM values of unique blocks according to the formula
\begin{equation}
\label{eq:bdm}
    \mathrm{BDM}(x) = \sum_{i} \left[ \mathrm{CTM}(y_i) + \log_2(m_i) \right],
\end{equation}
where $y_i$ ranges over the unique blocks of $x$ and $m_i$ is the number of occurrences of $y_i$. This approach is meant to enable tractable approximation of Kolmogorov complexity for arbitrarily large discrete datasets while remaining computationally feasible.

The decomposition itself can be carried out in different ways~\cite{zenil2016decomposition}. In the uncorrelated (default) variant, the string is split into non-overlapping blocks of size 12. In the correlated variant, the blocks are instead extracted with a sliding window of shift 1, so that overlapping blocks are counted and correlations across block boundaries are partially taken into account. Since every string of length $n$ then contributes $n - 11$ blocks rather than $\lfloor n/12 \rfloor$, the correlated BDM values span a considerably larger range, and the computation is correspondingly more expensive.

\section{Validation Against Exhaustive Enumeration}\label{app:validation}

\begin{figure}[ht]
    \centering
    \includegraphics[width=1.0\linewidth]{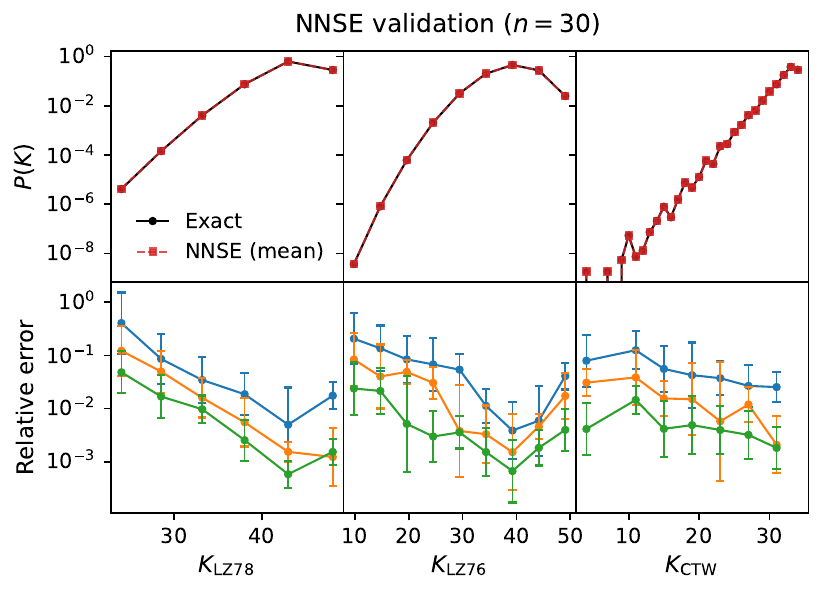}
    \caption{Validation of NNSE against exact complexity distributions at $n = 30$. Top: exact distribution (black) versus the mean NNSE estimate from ten independent $10^{6}$-step runs (red). Bottom: geometric mean pointwise relative error for $10^{4}$, $10^{5}$, and $10^{6}$ main steps (blue, orange, green), with error bars showing $\pm 1$ standard deviation across runs in log space. To avoid clutter, we only display every fourth point in the bottom right subplot for CTW errors.}
    \label{fig:mcmc_vs_exact}
\end{figure}

The NNSE method accurately reproduces exact complexity distributions where these can be computed. For $n = 30$, we enumerate all $2^{30} \approx 10^9$ binary strings, compute their LZ76, LZ78, and CTW complexities, and bin the values using the same thresholds as the NNSE runs as shown in Figure~\ref{fig:mcmc_vs_exact}. This gives the exact probability of each complexity level, directly comparable to the NNSE estimate bin by bin. We run NNSE for 3 main step sizes: $10^4, 10^5$ and $10^6$ with $1/10$ as many thermalization steps. For CTW, we now set the thresholds $K_i$ at intervals of $1$ and covering the entire complexity range. We also do $10$ identical runs for each step size to track the variation between the runs. We average over the runs and display error bars giving $\pm 1$ standard deviation across runs computed in log space (so that the bars are symmetric on the logarithmic axis).

The estimates track the exact distributions well across the entire range of complexity.  As expected, shorter runs show larger scatter.  The largest single-run deviations arise in the lowest-probability bins, whose exact mass can be as small as $10^{-8}$. This behavior is expected, since the innermost nested levels  accumulate the error of every conditional ratio in the nested product. The systematic improvement with increasing step count, together with the narrow spread across independent runs at fixed budget, indicates that residual error is dominated by Monte Carlo variance rather than bias. This agreement supports the reliability of NNSE at lengths where direct exhaustive enumeration is infeasible.

\section{LZ76 Symmetric}
\label{app:lz76_symmetric}

\begin{figure}
    \centering
    \includegraphics[width=1.0\linewidth]{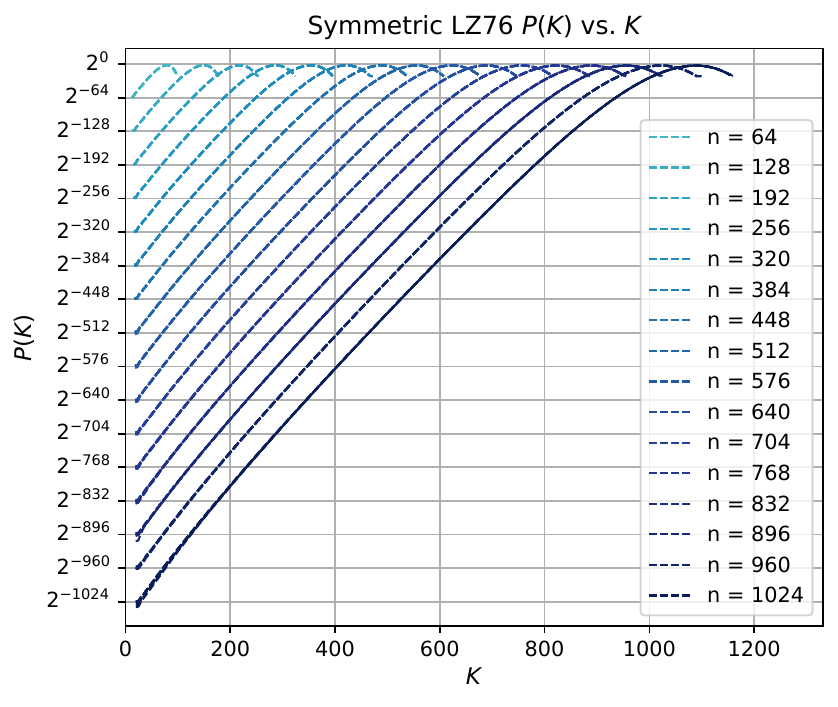}
    \caption{Complexity distribution for symmetric LZ76 obtained via NNSE with 8 runs per length, 10M main steps and 1M thermalization steps per run. Probability vs.\ symmetric LZ76 complexity, calculated using the NNSE method, for different string lengths $64 \leq n \leq 1024$ chosen at $\Delta n = 64$ increments. The  horizontal lines at $y = 2^{-n+1}$, correspond to the probability of the two simplest strings for length $n$. Each length contains 8 lines for independent runs, which helps illustrate statistical variation between runs.}
    \label{fig:LZ76_symmetric}
\end{figure}

We also studied symmetric LZ76 (SLZ)~\cite{dingle2018input} defined as 
\begin{equation}
    SLZ76(x_1 \dots x_n) = \frac12(LZ76(x_1 \dots x_n) + LZ76(x_n \dots x_1)),
\end{equation}
which has two nice properties as a complexity measure. First, symmetric LZ76 admits approximately twice as many complexity levels for every string, with $K$ values now coming in $\frac 12 \log_2 n$ intervals rather than just $\log_2 n$ for regular LZ76. Second, as it is symmetric, it is better suited for analyzing the sequences which do not have a clear direction in time~\cite{dingle2018input}. The NNSE method can estimate the complexity distribution for $n \leq 1024$ with little difference between different runs, as shown in Figure~\ref{fig:LZ76_symmetric}.  That NNSE is even better suited for symmetric LZ76 than regular LZ76 can be inferred from the fact that the variations between different runs are smaller.  The reasons may be due to the lower bit-flip sensitivity of the symmetrized version of LZ76.  

\section{More on CTW}
\label{app:ctw}

\begin{figure}
    \centering
    \includegraphics[width=1.0\linewidth]{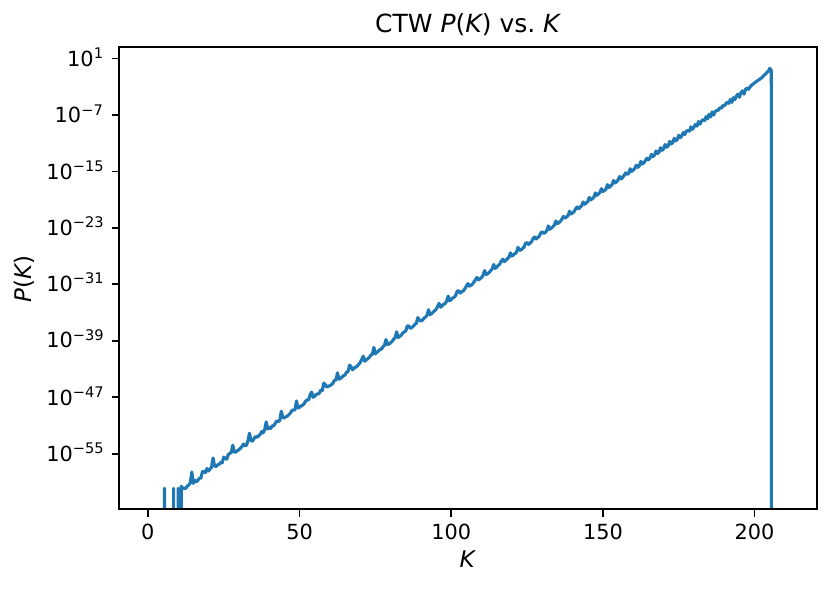}
    \caption{Complexity distribution for infinite-depth CTW obtained via NNSE for 10M main steps and 1M thermalization steps per run. Note the mostly linear relation between $K$ and $\log_2 P$ with small regularly spaced spikes across the entire range of $K$.}
    \label{fig:ctw_fine_grained}
\end{figure}

Since NNSE looks to be particularly well suited for CTW due to its low sensitivity to single-bit flips, we used NNSE to examine infinite-depth CTW in even more detail than the other complexity measures by separating the space into even more complexity levels $K_i$ at distance $\Delta K = K_{i+1} - K_i = 0.5$ apart rather than $\Delta K = \frac12 \log_2 n$ as in the earlier experiments, where $n = 200$. We looked at infinite-depth CTW, despite the increased computational cost, to avoid any possible patterns that could stem from choosing a particular depth. We did 4 identical runs of NNSE with 10M main steps and 1M thermalization steps per run. And the resulting $\log_2 P(K)$ values were all within $10\%$ of each other across the entire range of complexities up to typical strings with $K \approx 200$, with the variation being less than $1\%$ for most complexity values. We show the estimated distribution obtained from one run in Figure~\ref{fig:ctw_fine_grained}.

\begin{figure}
    \centering
    \includegraphics[width=1.0\linewidth]{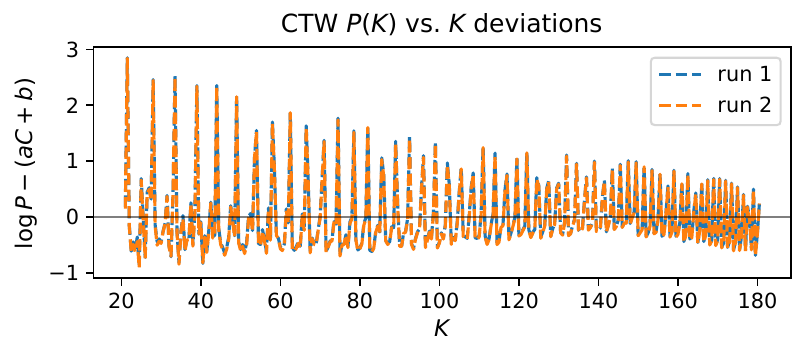}
    \caption{Deviations of infinite-depth CTW complexity distribution, obtained via NNSE, from the best-fit line show spikes at certain complexity values. Both the spacing between spikes and the magnitudes of the spikes seem to increase as $K$ goes to $0$. Two separate runs plotted to show agreement across different runs. }
    \label{fig:residuals_ctw}
\end{figure}

A linear relationship between $K$ and $\log_2 P$ was already expected from earlier results, but we also noticed regularly spaced spikes across the entire range of $K$. The spacing between the spikes as well as their magnitudes were consistent across runs as shown in Figure~\ref{fig:residuals_ctw}. This is also consistent with the spikes seen for CTW for the very short string length $n = 30$ obtained via exhaustive enumeration as shown in Figure~\ref{fig:mcmc_vs_exact}, supporting our belief that these spikes are not a result of experimental error resulting from NNSE, although we do not know the reason for these spikes. That NNSE is able to uncover such fine-grained patterns illustrates the usefulness of this method for studying complexity measures. 

\begin{figure}
    \centering
    \includegraphics[width=1.0\linewidth]{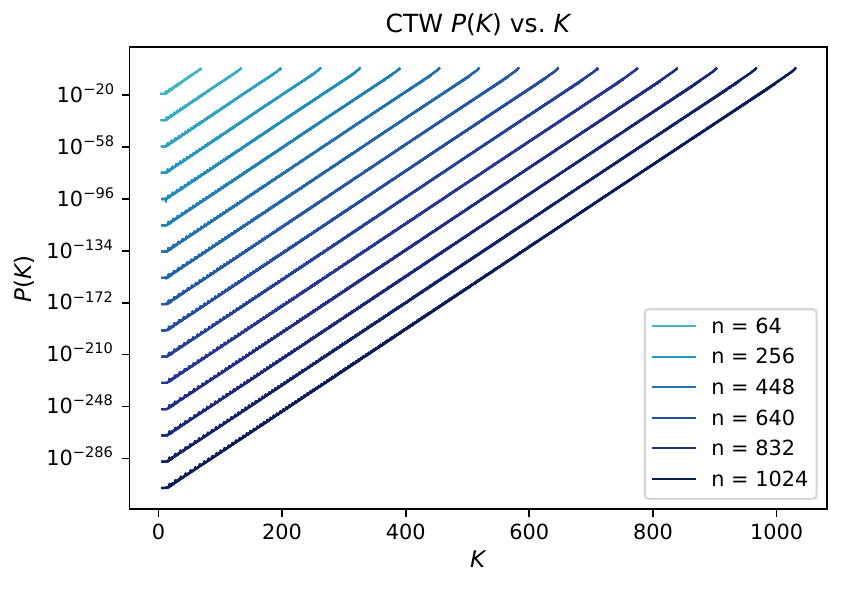}
    \caption{Complexity distribution for infinite-depth CTW obtained via NNSE for 1M main steps and 100K thermalization steps per run for $n = 64, 128, \dots, 1024$. Values with $P(K) = 0$ are omitted. 4 runs plotted for each $n$ but the different runs agree so well there's no visible deviation between them. Note the mostly linear relation between $K$ and $\log_2 P$ with small regularly spaced spikes across the entire range of $K$.}
    \label{fig:ctw_all}
\end{figure}

\begin{figure}
    \centering
    \includegraphics[width=1.0\linewidth]{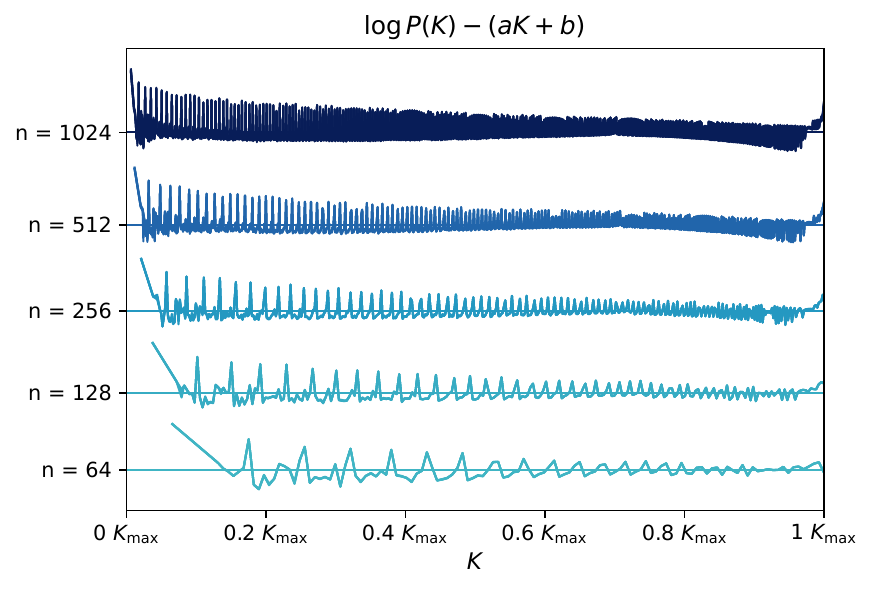}
    \caption{Deviations of infinite-depth CTW complexity distribution, obtained via NNSE, from the best-fit line at selected $n$ values. The $P(K)$ values are scaled to show the spikes clearly. The number of spikes seems to increase proportionally to $n$ with the spacing between them being roughly constant with respect to normalized $K$.}
    \label{fig:ctw_residuals}
\end{figure}

In addition, we also did 4 NNSE runs of 1M steps each for $n = 64, 128, \dots, 1024$ with $\Delta K = 0.5$ for infinite-depth CTW. We observed that both the linear relationship between $K$ and $P(K)$ and the spikes in $P(K)$ were consistent across different lengths and different runs agreed with each other very well despite the small number of main steps as shown in Figure~\ref{fig:ctw_all} and Figure~\ref{fig:ctw_residuals}. Since CTW has very low UCW sensitivity, NNSE appears to converge very quickly even for large $n$ values.

\begin{figure}
    \centering
    \includegraphics[width=1.0\linewidth]{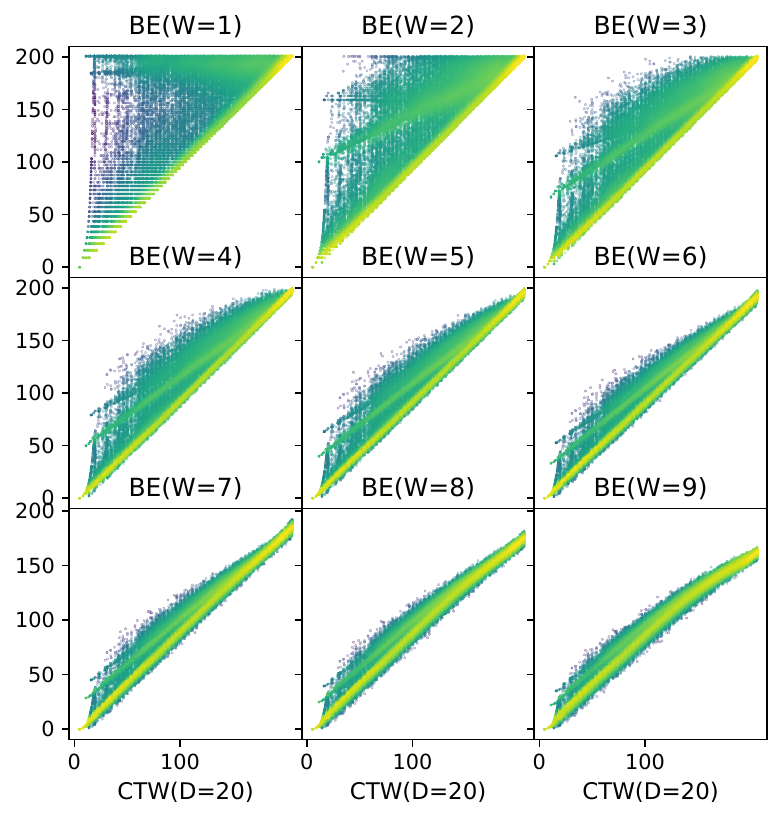}
    \caption{CTW vs.\ block entropy with small widths $W$ for length $n=200$ strings sampled via the NNSE method, with a brighter color indicating higher density. Here, the x values are CTW complexity and y values are block entropy. These plots reveal a number of patterns, including a concentration along the $x=y$ axis, where both measures give the same values, and secondary lines at higher values of block entropy, where there is a strong correlation. 
    The density of points is on a logarithmic scale. 
    We observe peaks in the scatter plots. We use logarithmic intensity to make the peaks more apparent. x values are CTW complexity and y values are block entropy.}
    \label{fig:ctw_be_plots}
\end{figure}

We also investigated CTW's relation to block entropy in more detail. One might expect that the block entropies of strings at different CTW complexities have a relatively uniform distribution. But we observe that strings at different complexities are much more likely to have some values of block entropy than other values. We find unexpected peaks in the scatter plots in Figure \ref{fig:ctw_be_plots}. The $x = y$ line is clearly visible but there are also additional $y = ax + b$ lines with $a = 1/2, 1/3$, or $1/4$ depending on the width of the blocks. The existence of such peaks, to our knowledge, is a novel finding, and is also unique to CTW (these lines are not apparent for LZ76/78). This shows that NNSE not only shows correlations but also fine-grained information for each individual complexity level. These additional lines would be completely invisible if we had sampled strings from the typical range of high complexity or even if we had found only the typical strings at each complexity level.

\section{Appendix: BDM $P_K$ distribution}\label{app:bdm}

\begin{figure}
    \centering
    \includegraphics[width=1.0\linewidth]{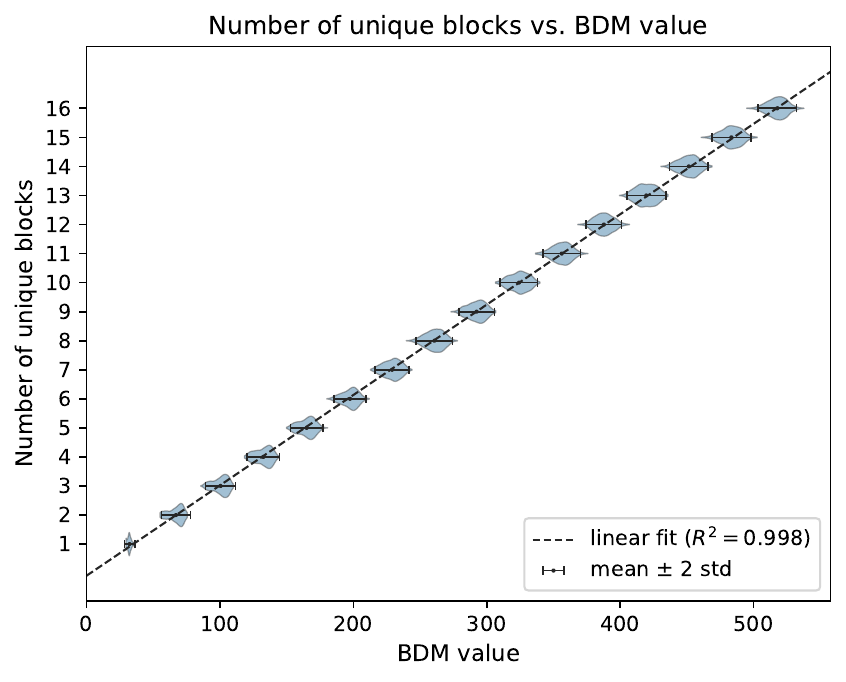}
    \caption{BDM complexity versus the number of unique 12-bit blocks for $N=10^4$ length-$200$ strings sampled from an NNSE run with uncorrelated BDM. Horizontal violins show the BDM distribution at each unique-block count. Points mark the mean BDM; whiskers are $\pm 2$ standard deviations. The dashed line is an ordinary least-squares fit of unique-block count against BDM, with the coefficient of determination $R^2$ given in the legend.}
    \label{fig:bdm_blocks}
\end{figure}

\begin{figure}
    \centering
    \includegraphics[width=1.0\linewidth]{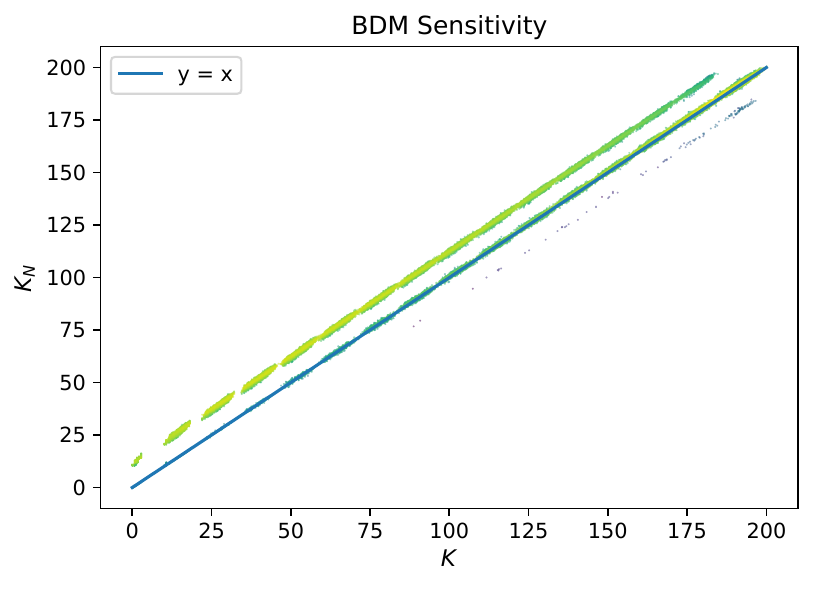}
    \caption{Scatter plot of complexity of a string upon a single bit-flip ($K_N$) vs.\ its original complexity $K$ for BDM (uncorrelated) for strings of length $n=200$. First, $100{,}000$ strings are sampled UCW from an NNSE run with 1 billion main steps. Then a single bit is flipped for all strings to obtain a neighboring string, one Hamming distance apart. The new complexity $K_N$ is plotted on the y-axis, and the original complexity $K$ on the x-axis. The three linear lines are a result of BDM's wobbly behavior and the small number of points with $K_N$ being one level below $K$ indicates that it's hard for NNSE to find strings with low complexity and converge to the true distribution $P(K)$. }
    \label{fig:bdm_sensitivity}
\end{figure}

What explains the bumps seen in Figure~\ref{fig:bdm} for uncorrelated BDM? We believe it's a result of how BDM splits a string into blocks of size $12$. Looking at equation~\ref{eq:bdm}, one can see that when there are $10$ blocks, the increase in complexity as a result of repeating a block is far less than that of adding a unique block as CTM values range from around $25.6$ to $37.5$ for length $12$ binary strings, which is far greater than the $\log_2$ values of small integers up to $16$. A string of length $200$ can be split into at most $16$ non-overlapping blocks of size exactly $12$ and thus, there can be anywhere from $1$ to $16$ unique CTM values for length $12$ substrings. This is consistent with the $15$ bumps in Figure~\ref{fig:bdm} plus the one flat bump between $K \approx 0 - 12$ (normalized value). We use NNSE to UCW sample $10,000$ strings and count the number of unique non-overlapping length $12$ substrings, as shown in Figure~\ref{fig:bdm_blocks}. We find a very strong correlation between the number of unique blocks and BDM complexity ($R^2 = 0.998$).  Oscillatory behavior can also be reproduced in  many other ways,  for example, sliding a fixed pattern through a string and measuring $K$ with BDM.    

This "bumpy" behavior helps explain why NNSE is  harder to converge for BDM than for some other measures: Bumps are hard to escape via single bit flips. Escaping each bump requires large jumps and the proportion of strings with neighbors that have lower complexity is very low, as shown in Figure~\ref{fig:bdm_sensitivity}.

\section{Appendix: Convergence and analysis of NNSE}\label{app:analysis}

Let $n\geq 1$ and $m\geq 0$ be integers, and suppose
\[
\varnothing\neq V_0\subseteq\cdots\subseteq V_m=\{0,1\}^n.
\]
We consider fully populated states. Define
$\Omega=V_0\times\cdots\times V_m$, and let $\mu$ be the uniform
distribution on $\Omega$, so $\mu(x)=|\Omega|^{-1}$ for $x\in\Omega$.

We write $P$ for the transition matrix of the NNSE on the state space $V_0 \times \dots \times V_m$ and ignore the mutation-skip-probability for now, as in~\cite{jorg2008neutral}. We also assume that all the $x_i$ have been filled or were initialized from $V_i$. In our algorithm, only the top level is filled initially, but through the mutation and swap steps, eventually all levels are filled. Each time step first mutates every $x_i$ and then runs the exchange sweep (Algorithm~\ref{alg:exchange}), so $P$ factorizes as
\begin{equation}
P = MS, \qquad M = M_0 \otimes \dots \otimes M_m, \qquad S = S_1 S_2 \cdots S_m,
\end{equation}
where $M_i$ is the stochastic matrix of the mutation $x' \gets \mathcal M(x_i)$ on $V_i$ and $S_i$ is the permutation matrix that swaps $x_{i-1}$ and $x_i$ whenever $x_i \in V_{i-1}$. We extend each factor to the full state space by letting it act as the identity on the other coordinates. The sweep visits $i = 1, \dots, m$ in increasing order, so $S_1$ comes first in the product.

Both $S_i$ and $M_i$ are symmetric. For a single bit flip, $M_i(x, x') = 1/n$ whenever $x, x' \in V_i$ differ in one bit, and $M_i(x, x)$ is the fraction of bit flips of $x$ that land in $V_m \backslash V_i$ and are therefore rejected. All other entries are zero, so $M_i$ is symmetric. The swap $S_i$ is an involution: applying it twice returns $(x_{i-1}, x_i)$ to where it started. Its matrix has $1$ on the diagonal for every state with $x_i \in V_i \backslash V_{i-1}$, has entry $1$ at $\big((x, y), (y, x)\big)$ for every $x, y \in V_{i-1}$, and is zero elsewhere, so it is symmetric too.

Each factor is therefore doubly stochastic: its rows and columns sum to one. Their product $P$ is also doubly stochastic, so $\mu P=\mu$. Thus the uniform distribution is stationary for the original NNSE. We cannot prove that this transition matrix is ergodic and irreducible for arbitrary nested sets $V_i$ because it always attempts to flip a single bit in each $x_i$. For some spaces, this can lead to periodic behavior and break the connectedness of $\Omega$ under $P$. Thus for the rest of this section, we analyze NNSE with a small change to the mutation step compared to the original paper~\cite{jorg2008neutral}. 

\begin{figure}
    \centering
    \includegraphics[width=\linewidth]{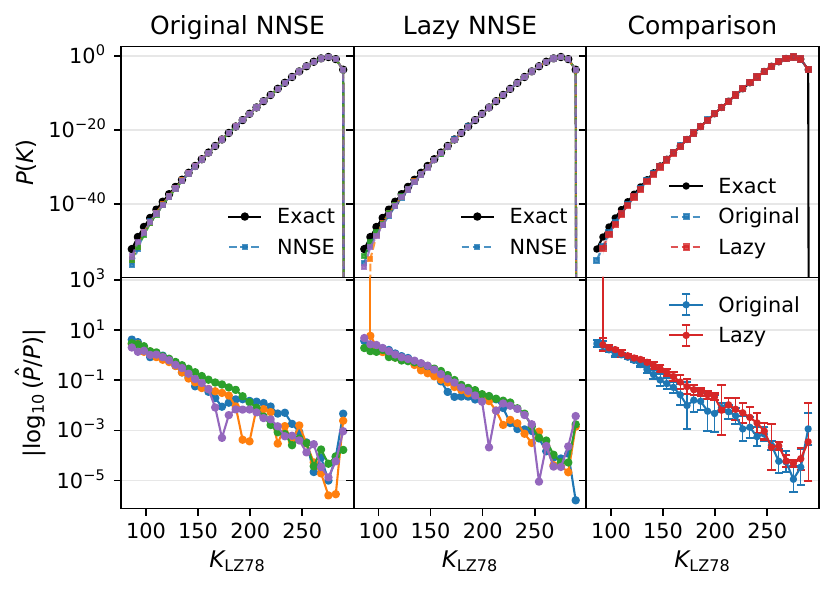}
    \caption{Comparison of the original NNSE vs. lazy NNSE with mutation-skip-probability $\eta = 1/2$. Displayed in the left and center columns are the $P(K_\text{LZ78})$ estimates, and their errors, for four runs per each variant of NNSE, with 10M thermalization steps and 100M main steps. In the right column, we display their mean estimates for $P(K_\text{LZ78})$ and their mean errors. The original NNSE variant ($\eta = 0$) clearly has a lower error rate across the entire complexity range. Note that since all the runs underestimate $P(K_\text{LZ78})$ at low $K_\text{LZ78}$, we used $\log |\hat P/P|$ rather than $|\hat P/P - 1|$ to measure error and show the underestimation of $\hat P$ more clearly. Since one lazy run estimated the lowest complexity's $P(K_\text{LZ78})$ to be $0$, that line goes to infinity in the plots}
    \label{fig:lazy_vs_nonlazy}
\end{figure}

This makes it possible to prove ergodicity, but in our experiments, we notice no meaningful difference between the two methods, except that the modified NNSE, not surprisingly, takes slightly longer to converge, as shown in Figure~\ref{fig:lazy_vs_nonlazy}, where we compared both methods for LZ78, the complexity measure  where NNSE struggled the most. In practice we use NNSE with either a very low probability of the lazy step, or even the original version in the limit of no lazy steps to maximise the efficiency.   In future work it will be interesting to find systems (that are not contrived)  for which the lazy variant provides better converged outcomes. 

For the rest of this section, we assume a lazy NNSE algorithm, i.e.\ $0<\eta<1$. At every time step, each level independently skips its mutation with probability $\eta$ and otherwise makes the original proposal. The exchange sweep is unchanged. We continue to write $M_i$, $M$ and $P$ for the resulting lazy transition matrices. Each factor remains doubly stochastic, so $\mu$ remains stationary.

\begin{theorem}
Lazy NNSE is irreducible and aperiodic, hence ergodic, on $\Omega$. Its unique stationary distribution is $\mu$, and
\begin{equation}
\lim_{t\to\infty}P^t(x,y)=\mu(y),\qquad x,y\in\Omega.
\end{equation}
\end{theorem}
\begin{proof}
With positive probability all mutations are skipped, so a step performs only the exchange sweep. Since the sweep is a permutation of the finite set $\Omega$, its inverse can also be performed by repeating such steps.

We can therefore mutate only $x_m$ and then undo the subsequent sweep. Since $V_m$ is connected under single-bit flips, this allows $x_m$ to be changed to any chosen element of $V_m$, with all lower coordinates restored at the end. Both this change and the sweep can be reversed by positive-probability paths.

Fix $a\in V_0$ and define $\bar a=(a,\ldots,a)$. Starting from any population, repeatedly set $x_m=a$ and perform a sweep. After $j$ repetitions,
\[
x_{m-j}=\cdots=x_{m-1}=a,\qquad 1\leq j\leq m.
\]
Indeed, the increasing-index sweep moves the existing copies down one level and places the new copy at level $m-1$. All these swaps are accepted because $a$ belongs to every level. After $m$ repetitions, setting $x_m=a$ once more gives $\bar a$. Since the construction can be reversed, every state communicates with $\bar a$, proving irreducibility.

At $\bar a$, skipping all mutations leaves the population unchanged even after the sweep. Hence $P(\bar a,\bar a)\geq\eta^{m+1}>0$, proving aperiodicity. Uniqueness of the stationary distribution and convergence to it now follow from the finite-state convergence theorem \cite[Sec.~2.1]{aldous-fill-2014}.
\end{proof}

Since $\Omega$ is finite, this implies a finite mixing time $\gamma$. The theorem above also implies that starting from any seed $(\varnothing, \varnothing, \dots, \varnothing, x_m)$ with $x_i$ unfilled for $i < m$, we will fill all levels $i$ eventually. Suppose we have only filled levels $i$ to $m$ up until some time $t$ but $x_{i-1} = \varnothing$. By the theorem above, $P$ is also ergodic on just the space $V_i \times \dots \times V_m$. Thus, after some finite amount of time, $x_i$ will be some element of $V_{i-1}$ since we assume all nested sets are nonempty. At that point, the swap operation will just copy $x_i$ to $x_{i-1}$ and thus all levels $i-1$ to $m$ will be filled. Thus, by induction, all levels $0$ to $m$ will eventually be filled.

\subsection{Accuracy of the size estimates}

Assume $m\geq1$, since for $m=0$ the only set size is already known. Write $N_i=|V_i|$, so $N_m=2^n$. Let $X^{(t)}=(X_0^{(t)},\ldots,X_m^{(t)})$ be the population after $t$ iterations, starting from any fully populated state. For $0\leq i<m$ and $x=(x_0,\ldots,x_m)\in\Omega$, define
\begin{equation}
 r_i=\frac{N_i}{N_{i+1}},
 \qquad f_i(x)=\mathbf 1[x_{i+1}\in V_i].
\end{equation}
After $T_0$ thermalization iterations, retain every subsequent iteration and estimate
\begin{equation}\label{eq:r_i_N_i_def}
\begin{aligned}
 \widehat r_i
 &=\frac1T\sum_{t=1}^{T}f_i(X^{(T_0+t)}),\qquad i<m,\\
 \widehat N_i
 &=2^n\prod_{j=i}^{m-1}\widehat r_j,\qquad 0\leq i\leq m.
\end{aligned}
\end{equation}
The empty product is one, so $\widehat N_m=N_m$. These are the NNSE estimates $\widehat N_i=2^n\widehat\pi_i$. Define their maximum relative error by
\begin{equation}\label{eq:E_def}
 \mathcal E=\max_{0\leq i\leq m}
 \left|\frac{\widehat N_i}{N_i}-1\right|.
\end{equation}

We write $\Pr_\mu$ and $\mathbb E_\mu$ for probability and expectation when the chain starts from its stationary distribution $\mu$. Under $\mu$, coordinate $i+1$ is uniform on $V_{i+1}$, so $\mathbb E_\mu f_i(X^{(t)})=r_i$. The ergodic theorem therefore gives $\widehat r_i\to r_i$ and $\widehat N_i\to N_i$ almost surely as $T\to\infty$, for any fixed $T_0$ \cite[Thm.~2.1]{aldous-fill-2014}. Since $0\leq\widehat N_i\leq2^n$ and $N_i\geq1$, we have $0\leq\mathcal E\leq2^n$. Bounded convergence also gives $\mathbb E\mathcal E\to0$.

Suppose that, for a fixed constant $a\geq0$,
\begin{equation}
 r_i\geq n^{-a},\qquad 0\leq i<m.
\end{equation}
For the complexity measures we looked at, we do not know a priori if there is a nontrivial $a$ which bounds $r_i$ but when picking complexity values $K_i$ that are close together, we observe that the ratios $r_i$ are not too low when using NNSE to estimate their sizes. Assuming $P(K) \propto 2^K$ and the separation between complexity values is $K_{i+1} - K_i \leq \log_2 n$, we would expect that $r_i \geq 1/n$. This motivates our stated choice of assumption but does not constitute a proof. 
The observed linear relationship between $\log_2 P(K)$ and $K$ in the NNSE runs can justify this assumption a posteriori for different complexity measures but also does not constitute a proof.

For two probability distributions $\nu,\rho$ on $\Omega$, define their total variation distance by
\[
 \|\nu-\rho\|_{\mathrm{TV}}
 =\frac12\sum_{x\in\Omega}|\nu(x)-\rho(x)|.
\]
Define
\begin{equation}
\begin{aligned}
 d(t)&=\max_{x\in\Omega}\|P^t(x,\cdot)-\mu\|_{\mathrm{TV}},\\
 \gamma&=\min\{t\geq1:d(t)\leq1/8\}.
\end{aligned}
\end{equation}
Thus $\gamma$ is the mixing time of the full population chain, measured in complete NNSE iterations. Submultiplicativity of $2d(t)$ gives
\begin{equation}
\label{eq:nnse-tv-decay}
 d(t)\leq2^{-\lfloor t/\gamma\rfloor}
 \qquad\text{for integers }t\geq0
\end{equation}
\cite[Lem.~2.20]{aldous-fill-2014}.

\begin{theorem}
\label{thm:nnse-size-accuracy}
For $0<\varepsilon<1$, the estimates satisfy
\begin{equation}
 \mathbb E\mathcal E\leq\varepsilon
\end{equation}
from any fully populated initial state using a thermalization and sample budget of orders
\begin{equation}
\begin{aligned}
 T_0&=O\!\left(\gamma\left[n+\log\frac1\varepsilon\right]\right),\\
 T&=O\!\left(\frac{\gamma m^2n^a}{\varepsilon^2}\right),
\end{aligned}
\end{equation}
with sufficiently large universal constants.
\end{theorem}
\begin{proof}
First suppose the samples are stationary. Define $U_i=|\widehat r_i/r_i-1|$ for $i<m$ and put $b=T/(72\gamma n^a)$. The Markov-chain Chernoff-Hoeffding bound \cite[Thm.~3]{chung2012chernoff} gives a universal constant $A>0$ such that
\begin{equation}
\label{eq:nnse-ratio-tail}
 \Pr_\mu(U_i \geq u)
 \leq A\exp\!\left[-b\min\{u^2,u\}\right],
 \qquad u\geq0.
\end{equation}
This bound applies to the full, possibly nonreversible chain and uses $r_i\geq n^{-a}$ to bound the mean of $U_i$.

Expanding $\mathcal E$ using \eqref{eq:E_def} and \eqref{eq:r_i_N_i_def} and applying the triangle inequality gives
\begin{equation}
\begin{aligned}
 \mathcal E
 &\leq\prod_{i=0}^{m-1}(1+U_i)-1\\
 &\leq\exp\!\left(\sum_{i=0}^{m-1}U_i\right)-1\\
 &\leq\frac1m\sum_{i=0}^{m-1}(e^{mU_i}-1),
\end{aligned}
\end{equation}
where the last inequality is convexity of the exponential. In particular, no independence between the ratio estimates is needed.

For $b\geq2m^2$, integration of \eqref{eq:nnse-ratio-tail} gives
\begin{equation}
\begin{aligned}
 \mathbb E_\mu(e^{mU_i}-1)
 &=m\int_0^\infty e^{mu}\Pr_\mu(U_i \geq u)\,du\\
 &\leq Am\left[
 e^{m^2/(4b)}\sqrt{\frac\pi b}
 +\frac{e^{m-b}}{b-m}\right]\\
 &=O\!\left(\frac m{\sqrt b}\right).
\end{aligned}
\end{equation}
Here we split the integral at $u=1$, complete the square in the first part, and integrate the exponential in the second part. Therefore,
\begin{equation}
\label{eq:nnse-stationary-error}
 \mathbb E_\mu\mathcal E
 =O\!\left(m\sqrt{\frac{\gamma n^a}{T}}\right).
\end{equation}
The stated sample budget $T$ makes this at most $\varepsilon/2$.

For an arbitrary initial state, choose $T_0$ so that $d(T_0)\leq\varepsilon/2^{n+1}$, using \eqref{eq:nnse-tv-decay}. Generating the next $T$ samples cannot increase total variation, so their joint distribution is within $d(T_0)$ of that obtained under stationarity. Since $0\leq\mathcal E\leq2^n$, it follows that
\begin{equation}
 \mathbb E\mathcal E
 \leq\mathbb E_\mu\mathcal E+2^n d(T_0)
 \leq\varepsilon.
\end{equation}

This proves that the stated choices of $T_0$ and $T$ ensure $\mathbb E \mathcal E$. The resulting total iteration budget is
\begin{equation}
 T + T_0 = O\!\left(\gamma\left[
 n+\log\frac1\varepsilon+
 \frac{m^2n^a}{\varepsilon^2}\right]\right).
\end{equation}
\end{proof}

\bibliographystyle{IEEEtran}
\bibliography{cited.bib}

\begin{thebibliography}{10}
\providecommand{\url}[1]{#1}
\csname url@samestyle\endcsname
\providecommand{\newblock}{\relax}
\providecommand{\bibinfo}[2]{#2}
\providecommand{\BIBentrySTDinterwordspacing}{\spaceskip=0pt\relax}
\providecommand{\BIBentryALTinterwordstretchfactor}{4}
\providecommand{\BIBentryALTinterwordspacing}{\spaceskip=\fontdimen2\font plus
\BIBentryALTinterwordstretchfactor\fontdimen3\font minus \fontdimen4\font\relax}
\providecommand{\BIBforeignlanguage}[2]{{%
\expandafter\ifx\csname l@#1\endcsname\relax
\typeout{** WARNING: IEEEtran.bst: No hyphenation pattern has been}%
\typeout{** loaded for the language `#1'. Using the pattern for}%
\typeout{** the default language instead.}%
\else
\language=\csname l@#1\endcsname
\fi
#2}}
\providecommand{\BIBdecl}{\relax}
\BIBdecl

\bibitem{ziv1978compression}
J.~Ziv and A.~Lempel, ``Compression of individual sequences via variable-rate coding,'' \emph{IEEE transactions on Information Theory}, vol.~24, no.~5, pp. 530--536, 1978.

\bibitem{doganaksoy2006lempel}
A.~Do{\u{g}}anaksoy and F.~G{\"o}lo{\u{g}}lu, ``On {L}empel-{Z}iv complexity of sequences,'' in \emph{Sequences and Their Applications--SETA 2006}, ser. Lecture Notes in Computer Science, vol. 4086.\hskip 1em plus 0.5em minus 0.4em\relax Berlin, Heidelberg: Springer, 2006, pp. 180--189.

\bibitem{dingle2018input}
K.~Dingle, C.~Q. Camargo, and A.~A. Louis, ``Input--output maps are strongly biased towards simple outputs,'' \emph{Nature communications}, vol.~9, no.~1, p. 761, 2018.

\bibitem{dingle2020generic}
K.~Dingle, G.~V. P{\'e}rez, and A.~A. Louis, ``Generic predictions of output probability based on complexities of inputs and outputs,'' \emph{Scientific reports}, vol.~10, no.~1, pp. 1--9, 2020.

\bibitem{mingard2019neural}
C.~Mingard, J.~Skalse, G.~Valle-P{\'e}rez, D.~Mart{\'\i}nez-Rubio, V.~Mikulik, and A.~A. Louis, ``Neural networks are a priori biased towards boolean functions with low entropy,'' \emph{arXiv preprint arXiv:1909.11522}, 2019.

\bibitem{johnston2022symmetry}
I.~G. Johnston, K.~Dingle, S.~F. Greenbury, C.~Q. Camargo, J.~P. Doye, S.~E. Ahnert, and A.~A. Louis, ``Symmetry and simplicity spontaneously emerge from the algorithmic nature of evolution,'' \emph{Proceedings of the National Academy of Sciences}, vol. 119, no.~11, p. e2113883119, 2022.

\bibitem{mingard2025deep}
C.~Mingard, H.~Rees, G.~Valle-P{\'e}rez, and A.~A. Louis, ``Deep neural networks have an inbuilt occam’s razor,'' \emph{Nature Communications}, vol.~16, no.~1, p. 220, 2025.

\bibitem{li2008introduction}
M.~Li and P.~Vitanyi, \emph{An introduction to Kolmogorov complexity and its applications}.\hskip 1em plus 0.5em minus 0.4em\relax Springer-Verlag New York Inc, 2008.

\bibitem{jorg2008neutral}
T.~Jorg, O.~Martin, and A.~Wagner, ``{Neutral network sizes of biological RNA molecules can be computed and are not atypically small},'' \emph{BMC bioinformatics}, vol.~9, no.~1, p. 464, 2008.

\bibitem{merhav1998universal}
N.~Merhav and M.~Feder, ``Universal prediction,'' \emph{IEEE Transactions on Information Theory}, vol.~44, no.~6, pp. 2124--2147, 1998.

\bibitem{merhav2019universal}
N.~Merhav and A.~Cohen, ``Universal randomized guessing with application to asynchronous decentralized brute--force attacks,'' \emph{IEEE Transactions on Information Theory}, vol.~66, no.~1, pp. 114--129, 2019.

\bibitem{lempel1976complexity}
A.~Lempel and J.~Ziv, ``On the complexity of finite sequences,'' \emph{IEEE Transactions on information theory}, vol.~22, no.~1, pp. 75--81, 1976.

\bibitem{willems1995context}
F.~M. Willems, Y.~M. Shtarkov, and T.~J. Tjalkens, ``The context-tree weighting method: Basic properties,'' \emph{IEEE transactions on information theory}, vol.~41, no.~3, pp. 653--664, 1995.

\bibitem{willems1998context}
F.~M.~J. Willems, ``The context-tree weighting method: Extensions,'' \emph{IEEE Transactions on Information Theory}, vol.~44, no.~2, pp. 792--798, 1998.

\bibitem{zenil2016decomposition}
H.~Zenil, F.~Soler-Toscano, N.~A. Kiani, S.~Hern{\'a}ndez-Orozco, and A.~Rueda-Toicen, ``A decomposition method for global evaluation of {S}hannon entropy and local estimations of algorithmic complexity,'' \emph{arXiv preprint arXiv:1609.00110}, 2016.

\bibitem{cover2006elements}
T.~Cover and J.~Thomas, \emph{Elements of information theory}.\hskip 1em plus 0.5em minus 0.4em\relax John Wiley and Sons, 2006.

\bibitem{kaspar1987easily}
F.~Kaspar and H.~Schuster, ``Easily calculable measure for the complexity of spatiotemporal patterns,'' \emph{Physical Review A}, vol.~36, no.~2, pp. 842--848, 1987.

\bibitem{begleiter2004prediction}
R.~Begleiter, R.~El-Yaniv, and G.~Yona, ``On prediction using variable order markov models,'' \emph{Journal of Artificial Intelligence Research}, vol.~22, pp. 385--421, 2004.

\bibitem{skilling2006nested}
J.~Skilling, ``Nested sampling for general bayesian computation,'' \emph{Bayesian Analysis}, vol.~1, no.~4, pp. 833--859, 2006.

\bibitem{aldous-fill-2014}
D.~Aldous and J.~A. Fill, ``Reversible {Markov} chains and random walks on graphs,'' 2002, unfinished monograph, recompiled 2014, available at \url{http://www.stat.berkeley.edu/$\sim$aldous/RWG/book.html}.

\bibitem{chung2012chernoff}
\BIBentryALTinterwordspacing
K.-M. Chung, H.~Lam, Z.~Liu, and M.~Mitzenmacher, ``{Chernoff-Hoeffding} bounds for {Markov} chains: Generalized and simplified,'' in \emph{29th International Symposium on Theoretical Aspects of Computer Science (STACS 2012)}, ser. Leibniz International Proceedings in Informatics (LIPIcs), C.~D{\"u}rr and T.~Wilke, Eds., vol.~14.\hskip 1em plus 0.5em minus 0.4em\relax Dagstuhl, Germany: Schloss Dagstuhl--Leibniz-Zentrum f{\"u}r Informatik, 2012, pp. 124--135. [Online]. Available: \url{https://drops.dagstuhl.de/entities/document/10.4230/LIPIcs.STACS.2012.124}
\BIBentrySTDinterwordspacing

\bibitem{estevez2013nonrandomness}
E.~Estevez-Rams, R.~Lora~Serrano, B.~Arag{\'o}n~Fern{\'a}ndez, and I.~Brito~Reyes, ``On the non-randomness of maximum lempel ziv complexity sequences of finite size,'' \emph{Chaos: An Interdisciplinary Journal of Nonlinear Science}, vol.~23, no.~2, p. 023118, June 2013.

\bibitem{vitter1985random}
J.~S. Vitter, ``Random sampling with a reservoir,'' \emph{ACM Transactions on Mathematical Software}, vol.~11, no.~1, pp. 37--57, 1985.

\bibitem{jacquet1995asymptotic}
P.~Jacquet and W.~Szpankowski, ``Asymptotic behavior of the lempel-ziv parsing scheme and digital search trees,'' \emph{Theoretical Computer Science}, vol. 144, no. 1-2, pp. 161--197, 1995.

\bibitem{zenil2014correlation}
H.~Zenil, F.~Soler-Toscano, K.~Dingle, and A.~A. Louis, ``Correlation of automorphism group size and topological properties with program-size complexity evaluations of graphs and complex networks,'' \emph{Physica A: Statistical Mechanics and its Applications}, vol. 404, pp. 341--358, 2014.

\bibitem{soler2014calculating}
F.~Soler-Toscano, H.~Zenil, J.-P. Delahaye, and N.~Gauvrit, ``Calculating {K}olmogorov complexity from the output frequency distributions of small {T}uring machines,'' \emph{PLoS ONE}, vol.~9, no.~5, p. e96223, 2014.

\bibitem{balle2017endtoend}
\BIBentryALTinterwordspacing
J.~Ball{\'e}, V.~Laparra, and E.~P. Simoncelli, ``End-to-end optimized image compression,'' in \emph{International Conference on Learning Representations (ICLR)}, 2017. [Online]. Available: \url{https://arxiv.org/abs/1611.01704}
\BIBentrySTDinterwordspacing

\bibitem{mahoney2005adaptive}
\BIBentryALTinterwordspacing
M.~V. Mahoney, ``Adaptive weighing of context models for lossless data compression,'' Florida Institute of Technology, Department of Computer Science, Tech. Rep. CS-2005-16, 2005. [Online]. Available: \url{https://repository.fit.edu/ces_faculty/166/}
\BIBentrySTDinterwordspacing

\bibitem{blumer1985}
A.~Blumer, J.~Blumer, A.~Ehrenfeucht, D.~Haussler, M.~T. Chen, and J.~Seiferas, ``The smallest automaton recognizing the subwords of a text,'' \emph{Theoretical Computer Science}, vol.~40, pp. 31--55, 1985.

\bibitem{lz78_linear}
Y.~Nakashima, T.~I, S.~Inenaga, H.~Bannai, and M.~Takeda, ``Constructing {LZ78} tries and position heaps in linear time for large alphabets,'' \emph{Information Processing Letters}, vol. 115, no.~9, pp. 655--659, 2015.

\bibitem{veness2011monte}
J.~Veness, K.~S. Ng, M.~Hutter, W.~Uther, and D.~Silver, ``A monte-carlo {AIXI} approximation,'' \emph{Journal of Artificial Intelligence Research}, vol.~40, pp. 95--142, 2011.

\end{thebibliography}

\end{document}